\documentclass[aps,prl,reprint,twocolumn,superscriptaddress,floatfix,nofootinbib,longbibliography]{revtex4-2}
\usepackage{graphicx,amsmath,amsfonts,amssymb,amsthm,xr}
\usepackage{epsfig,amsmath,amssymb,color,dsfont,upgreek,physics}
\usepackage{mathrsfs}
\usepackage{mathtools}
\usepackage{bbold}
\usepackage{float}

\usepackage[bookmarks=true,colorlinks,linkcolor=OrangeRed,urlcolor=NavyBlue,citecolor=RoyalBlue]{hyperref}
\usepackage[dvipsnames]{xcolor}
\usepackage{orcidlink}

\usepackage{graphicx}
\usepackage{dcolumn}
\usepackage{bm}
\usepackage{color}

\definecolor{maroon}{RGB}{128, 0, 0}

\usepackage{orcidlink}
\newcommand{\orcidjulian}{\orcidlink{0000-0002-7011-6477}}
\newcommand{\orcidivano}{\orcidlink{0000-0001-5690-1981}}
\newcommand{\orcidguoxian}{\orcidlink{0000-0001-7936-762X}}
\newcommand{\orcidjesse}{\orcidlink{0000-0003-0415-0690}}
\newcommand{\orcidjad}{\orcidlink{0000-0002-0659-7990}}
\newcommand{\orcidanthony}{\orcidlink{0009-0004-3948-7239}}

\newcommand{\ibmquantum}{IBM Quantum, IBM Research Europe - Zurich, 8803 R\"uschlikon, Switzerland}
\newcommand{\epfl}{Institute of Physics, \'Ecole Polytechnique F\'ed\'erale de Lausanne (EPFL), CH-1015 Lausanne, Switzerland}
\newcommand{\ethzurich}{Institute for Theoretical Physics, ETH Z\"urich, 8093 Z\"urich, Switzerland}
\newcommand{\mpq}{Max Planck Institute of Quantum Optics, 85748 Garching, Germany}
\newcommand{\lmu}{Department of Physics and Arnold Sommerfeld Center for Theoretical Physics (ASC), Ludwig Maximilian University of Munich, 80333 Munich, Germany}
\newcommand{\mcqst}{Munich Center for Quantum Science and Technology (MCQST), 80799 Munich, Germany}
\newcommand{\mitt}{Research Laboratory of Electronics, MIT-Harvard Center for Ultracold Atoms,
Department of Physics, Massachusetts Institute of Technology, MA 02139, USA}

\begin{document}

\title{Observation of hadron scattering in a lattice gauge theory on a quantum computer}

\author{Julian Schuhmacher$^{\orcidjulian}$}
\affiliation{\ibmquantum}
\affiliation{\epfl}

\author{Guo-Xian Su$^{\orcidguoxian}$}
\affiliation{\mitt}

\author{Jesse J.~Osborne$^{\orcidjesse}$}
\affiliation{\mpq}
\affiliation{\mcqst}

\author{Anthony Gandon$^{\orcidanthony}$}
\affiliation{\ibmquantum}
\affiliation{\ethzurich}

\author{Jad C.~Halimeh$^{\orcidjad}$}
\email{jad.halimeh@physik.lmu.de}
\affiliation{\mpq}
\affiliation{\lmu}
\affiliation{\mcqst}

\author{Ivano Tavernelli$^{\orcidivano}$}
\email{ita@zurich.ibm.com}
\affiliation{\ibmquantum}

\maketitle

\textbf{Scattering experiments are at the heart of high-energy physics (HEP), breaking matter down to its fundamental constituents, probing its formation, and providing deep insight into the inner workings of nature~\cite{Ellis_book}. In the current huge drive to forge quantum computers into complementary venues that are ideally suited to capture snapshots of far-from-equilibrium HEP dynamics, a major goal is to utilize these devices for scattering experiments~\cite{Bauer_review,dimeglio2023quantum}. A major obstacle in this endeavor has been the hardware overhead required to access the late-time post-collision dynamics while implementing the underlying gauge symmetry. Here, we report on the first quantum simulation of scattering in a lattice gauge theory (LGT), performed on \texttt{IBM}'s \texttt{ibm\_marrakesh} quantum computer. Specifically, we quantum-simulate the collision dynamics of electrons and positrons as well as mesons in a $\mathrm{U}(1)$ LGT representing $1+1$D quantum electrodynamics (QED), uncovering rich post-collision dynamics that we can precisely tune with a topological $\Theta$-term and the fermionic mass. By monitoring the time evolution of the scattering processes, we are able to distinguish between two main regimes in the wake of the collision. The first is characterized by the delocalization of particles when the topological $\Theta$-term is weak, while the second regime shows localized particles with a clear signature when the $\Theta$-term is nontrivial. Furthermore, we show that by quenching to a small mass at the collision point, inelastic scattering occurs with a large production of matter reminiscent of quantum many-body scarring~\cite{Serbyn2020}. Our work provides a major step forward in the utility of quantum computers for investigating the real-time quantum dynamics of HEP collisions.}

Gauge theories are a fundamental framework underpinning the Standard Model of particle physics, providing an overarching description of interactions between elementary particles as mediated by gauge fields through gauge symmetries \cite{Weinberg_book,Zee_book,Gattringer_book}. The latter stipulate an intrinsic relationship between matter and gauge fields, with Gauss's law from QED being a prime example, coupling electrons and positrons to the electromagnetic field at each point in space. LGTs, the lattice formulation of gauge theories \cite{Kogut_review,Rothe_book}, were initially conceived to gain insight into quark confinement \cite{Wilson1974}, but have since been shown to be powerful tools for understanding various phenomena in HEP and even condensed matter and quantum information \cite{Wegner1971,wen2004quantum,Lee2008}.

\begin{figure*}[t!]
    \centering
    \includegraphics[width=0.95\textwidth]{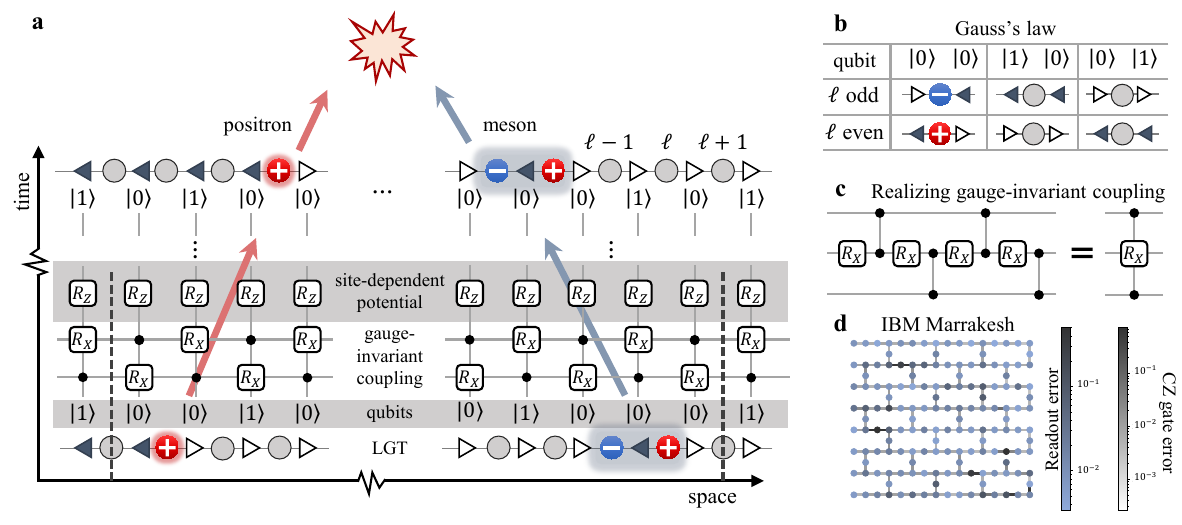}
    \caption{
        Particle scattering on a quantum computer.
        \textbf{a} An example of qubit representation of particles in a lattice gauge theory, and subsequent Quantum circuit to implement time evolution and particle scattering. Coupling is turned off for the initial time steps at the dashed lines to prepare moving wavepackets toward each other.
        \textbf{b} Gauss's law in the $\hat{G}_\ell\ket{\Psi}=0$ gauge sector of the spin-1/2 lattice QED model, and their qubit representation.
        \textbf{c} The quantum circuit implementing the gauge-invariant coupling term in the Hamiltonian, where dynamical gauge fields couples the hopping of matter fields.
        \textbf{d} The IBM Marrakesh quantum processor employed in this work. The color of the nodes represents the readout error of the corresponding qubits. The color of the links represents the CZ gate error between the connected qubits. (Reported error rates obtained on April 11, 2025).}
    \label{fig:quantum_circuit}
\end{figure*}

Scattering experiments are the basis for particle colliders such as the Large Hadron Collider (LHC) and the Relativistic Heavy Ion Collider (RHIC). They form our knowledge of subatomic structures, facilitate the discovery of new particles \cite{ATLAS2012,CMS2012}, and mimic the conditions of the early universe through the creation of quark-gluon plasmas \cite{Adcox2005,Back2005,Arsene2005}. So successful have such colliders been that works are in plan for a high-luminosity upgrade for the LHC and the Future Circular Collider at CERN, which will probe physics beyond the Standard Model \cite{FCC}. In a scattering process, particles travel toward and collide with each other, interact, and then produce a post-collision state. The quantum wave function encodes the entire history of this scattering process and provides information on how interactions between the colliding particles gave rise to their post-collision counterparts. 

Capturing snapshots of the real-time dynamics of scattering processes is a frontier theoretical endeavor. In this vein, quantum simulators of LGTs provide an ideal platform that enables first-principles studies of LGT dynamics \cite{Dalmonte_review, Pasquans_review, Zohar_review, aidelsburger2021cold, Zohar_NewReview, klco2021standard, Cheng_review, Halimeh_review, Cohen:2021imf, Lee:2024jnt, Turro:2024pxu}, allowing them to be potentially useful complementary venues to particle colliders \cite{Bauer_review, dimeglio2023quantum}. Recent years have witnessed a tremendous surge of LGT quantum simulation experiments \cite{Martinez2016,Klco2018,Goerg2019,Schweizer2019,Mil2020,Yang2020,Wang2021,Zhou2022,Wang2023,Zhang2023,Su2022,Ciavarella2024quantum,Ciavarella:2024lsp,de2024observationstringbreakingdynamicsquantum,liu2024stringbreakingmechanismlattice,Farrell:2023fgd,Farrell:2024fit,zhu2024probingfalsevacuumdecay,Ciavarella:2021nmj,Ciavarella:2023mfc,Ciavarella:2021lel,Gustafson:2023kvd,Gustafson:2024kym,Lamm:2024jnl,Farrell:2022wyt,Farrell:2022vyh,Li:2024lrl,Zemlevskiy:2024vxt,Lewis:2019wfx,Atas:2021ext,ARahman:2022tkr,Atas:2022dqm,Mendicelli:2022ntz,Kavaki:2024ijd,Than:2024zaj,Angelides2025first,alexandrou2025realizingstringbreakingdynamics,cochran2024visualizingdynamicschargesstrings,gyawali2024observationdisorderfreelocalizationefficient,gonzalezcuadra2024observationstringbreaking2,crippa2024analysisconfinementstring2,datla2025statisticallocalizationrydbergsimulator}, but a fully fledged quantum simulation of scattering dynamics in an LGT has been lacking until now. Whereas quantum simulations of scattering in non-LGT Hamiltonians have been performed   \cite{Chai2023fermionic,zemlevskiy2024scalablequantumsimulationsscattering,farrell2025digitalquantumsimulationsscattering,ingoldby2025realtimescatteringquantumcomputers}, only wave-packet preparation in LGTs has been carried out on quantum hardware \cite{Farrell2024hadrondynamics,Davoudi2024scatteringwavepreparation}. To faithfully probe HEP collisions like those at the LHC and RHIC, an LGT implementation is a necessary step in that direction.

In this work, we report on the first\footnote{See also the parallel submission by Z.~Davoudi, C.-C.~Hsieh, and S.~V.~Kadam in today's \texttt{arXiv} listing.} quantum simulation of scattering processes in an LGT, performed on \texttt{IBM}'s \texttt{ibm\_marrakesh} quantum computer; see Fig.~\ref{fig:quantum_circuit}. We focus on a $\mathrm{U}(1)$ LGT representing QED and quantum-simulate electron-positron and meson-meson collisions.
The quantum circuits implementing the gauge-invariant dynamics are obtained via a first-order Trotterization of the time-evolution operator of the LGT Hamiltonian in its spin representation, obtained by integrating out Gauss's law to enforce gauge invariance.
The nature of the dynamics, i.e., the particles that we are simulating, are encoded in gauge-invariant initial states, which can easily be prepared with a quantum circuit consisting of just one layer of single-qubit rotations. We apply this setup to perform quantum simulations of electron-positron and meson-meson scatterings on systems of up to $45$ qubits.
To obtain accurate results for up to $35$ Trotter steps (two-qubit gate depth of $280$), we apply an error mitigation strategy called marginal Distribution Error Mitigation (mDEM), which is an adaption of the recently introduced DEM~\cite{gonzales2025quantum} to the mitigation of marginal distributions and, therefore, to the mitigation of local expectation values.
We demonstrate that it is possible to faithfully follow the evolution of the particles far beyond their collision point.

\textbf{Model and quantum circuit.---}We consider the $1+1$D $\mathrm{U}(1)$ LGT given by the Hamiltonian
\begin{align}\nonumber
    \hat{H}_\text{LGT}=&{-}\frac{\kappa}{2}\sum_{\ell=1}^{L-1}\Big(\hat{\psi}_\ell^\dagger\hat{U}_{\ell,\ell+1}\hat{\psi}_{\ell+1}{+}\text{H.c.}\Big)\\\label{eq:U1LGT}
    &{+}m\sum_{\ell=1}^L(-1)^\ell\hat{\psi}_\ell^\dagger\hat{\psi}_\ell{+}\sum_{\ell=1}^{L-1}\Big(\hat{E}_{\ell,\ell+1}{+}\hat{E}_{\text{bg}}\Big)^2,
\end{align}
where the staggered fermionic operators $\hat{\psi}_\ell^{(\dagger)}$ represent the matter degrees of freedom at site $\ell$, $m$ is the fermionic mass, $L$ is the total number of sites, $\hat{U}_{\ell,\ell+1}$ is dynamical gauge field at the link between sites $\ell$ and $\ell+1$, which couples the gauge-invariant hopping of fermionic matter, and $\hat{E}_{\ell,\ell+1}$ is the electric field at the same link. The gauge-invariant hopping results in the change of electric fields whose energies are represented by the last term of the Hamiltonian. The external background electric field $\hat{E}_{\text{bg}}=g\Theta/(2\pi)$ is related to the topological $\Theta$-angle and provides a confining potential between electron-positron pairs. Adopting the quantum link model formulation \cite{Chandrasekharan1997,Wiese_review}, we represent the gauge and electric fields by the spin-$1/2$ operators: $\hat{U}_{\ell,\ell+1}=\hat{s}^+_{\ell,\ell+1},\,\hat{E}_{\ell,\ell+1}=g\hat{s}^z_{\ell,\ell+1}$, where $g$ is the gauge coupling.
The last term of the Hamiltonian~\eqref{eq:U1LGT} thus becomes $\chi\sum_{\ell}^{L-1}\hat{s}^z_{\ell,\ell+1}$, where $\chi=g^2(\Theta-\pi)/(2\pi)$ \cite{Halimeh2022tuning,Cheng2022}.
This model is known to exhibit rich dynamics \cite{Surace2020,Desaules2022weak,Desaules2022prominent,Desaules2024ergodicitybreaking} and has been used to investigate microscopic confinement dynamics \cite{Zhang2023}, false vacuum decay \cite{zhu2024probingfalsevacuumdecay}, and string breaking \cite{liu2024stringbreakingmechanismlattice} in a cold-atom quantum simulator. At $\chi=0$, it hosts a quantum phase transition related to the spontaneous breaking of charge conjugation and parity symmetry, with a quantum critical point $m_c{=}0.3275\kappa$ \cite{Coleman1976,Yang2020}.
The generator of the $\mathrm{U}(1)$ gauge symmetry of the Hamiltonian~\eqref{eq:U1LGT} is $\hat{G}_\ell{=}\hat{s}^z_{\ell,\ell+1}{-}\hat{s}^z_{\ell-1,\ell}{-}\hat{\psi}_\ell^\dagger\hat{\psi}_\ell{-}[({-}1)^\ell{-}1]/2$. 
We restrict the dynamics to the sector of Gauss's law, i.e., $\hat{G}_\ell\ket{\Psi}=0,\,\forall \ell$; see Fig.~\ref{fig:quantum_circuit}\textbf{b}. 
We employ Gauss's law to integrate out the matter fields, index all gauge sites by $j$ and obtain the spin model $\hat{H} = \sum_j\{\frac{\kappa}{2} \hat{P}_{j-1} \hat{\sigma}^x_j \hat{P}_{j+1}+[(-1)^j\chi-2m]\hat{N}_j\}$, where $\hat{P}_j = (\hat{\mathbb{I}} + \hat{\sigma}^z_j)/2 = \ketbra{0}$ is the projector to the ground state at site $j$, $\hat{N}_j = (\hat{\mathbb{I}} - \hat{\sigma}^z_j)/2 = \ketbra{1}$ is the projector to the excited state, and $\hat{\sigma}^\alpha \in \{\hat{\sigma}^x, \hat{\sigma}^y, \hat{\sigma}^z\}$ are the Pauli matrices \cite{Surace2020}.

With the matter fields integrated out, the two vacuum states $\ket{\ldots\triangleright\triangleright\triangleright\triangleright\ldots}$ and $\ket{\ldots\blacktriangleleft\blacktriangleleft\blacktriangleleft\blacktriangleleft\ldots}$ thus map to the two antiferromagnetically ordered states in the spin model, represented as $\ket{\ldots0101\ldots}$ and $\ket{\ldots1010\ldots}$ on the quantum computer. The electrons (positrons) are therefore encoded as domain walls. For a spin chain indexed starting at $1$, state $\ket{010010\ldots}$ represents a single electron $\ket{\triangleright\triangleright\triangleright e^-\blacktriangleleft\blacktriangleleft\blacktriangleleft\ldots}$, while state $\ket{100101\ldots}$ represents a single positron $\ket{\blacktriangleleft\blacktriangleleft e^+\triangleright\triangleright\triangleright\triangleright\ldots}$. As such, $\ket{0100010\ldots}$ is the meson state $\ket{\triangleright\triangleright\triangleright e^-\blacktriangleleft e^+\triangleright\triangleright\triangleright\ldots}$, and $\ket{10001\ldots}$ is the antimeson state $\ket{\blacktriangleleft\blacktriangleleft e^+\triangleright e^-\blacktriangleleft\blacktriangleleft\ldots}$. All these product states can be directly initialized on the quantum computer.

\begin{figure}[t!]
    \centering
    \includegraphics[width=\columnwidth]{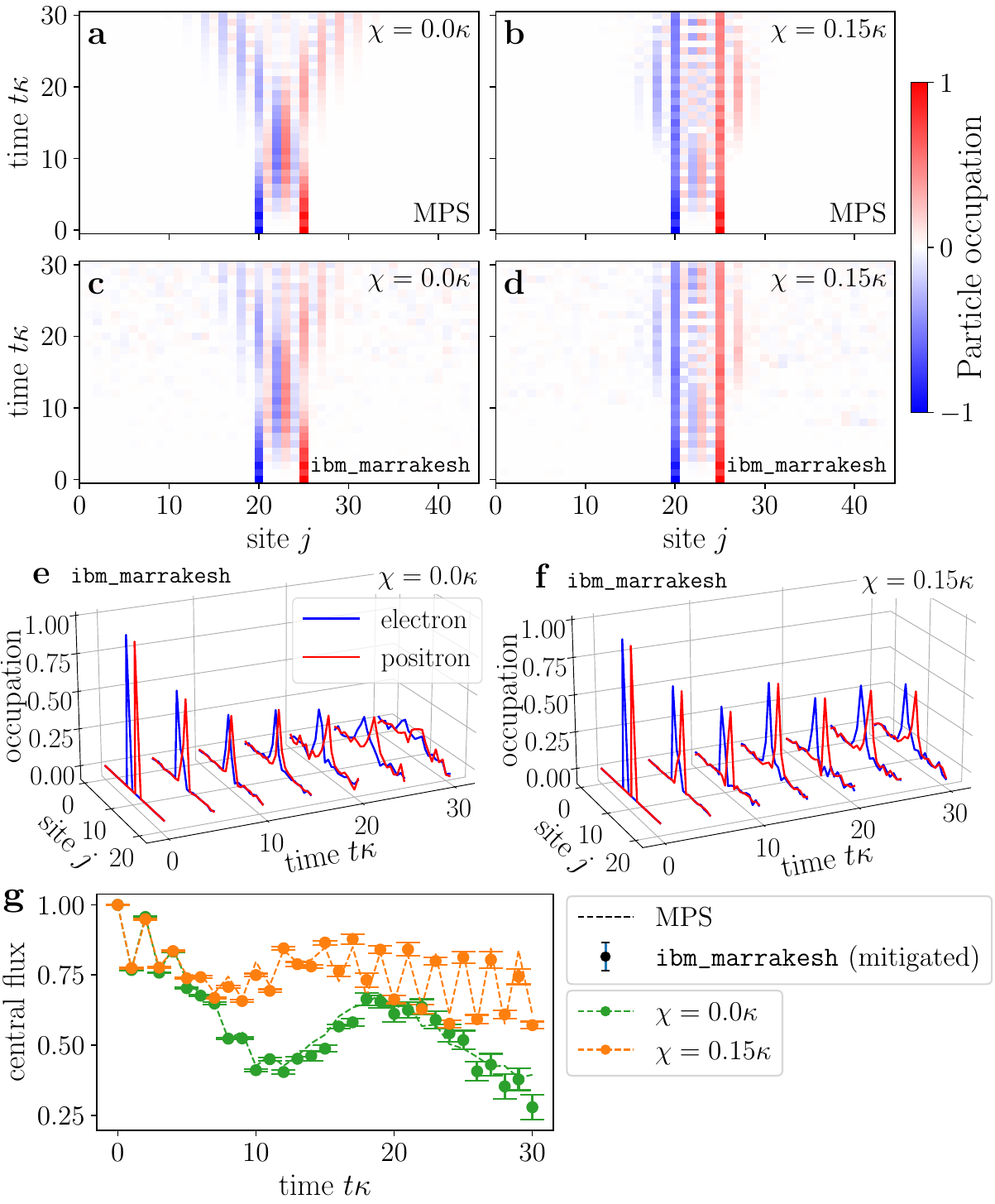}
    \caption{Dynamics of electron-positron scattering without and with confinement for electrons and positrons with mass $m = 1.5\kappa$. \textbf{a},\textbf{c},\textbf{e} Evolution of the particle occupation number during the scattering of an electron and a positron. \textbf{a} Classical reference obtained with MPS simulations. \textbf{c} Mitigated results obtained on quantum hardware. \textbf{e} Different representation of the same data as in \textbf{c}. Electron occupation number is shown in blue, positron occupation number in red. \textbf{b},\textbf{d},\textbf{f} Occupation number for the electron-positron scattering with a confining potential of $\chi = 0.15 \kappa$. \textbf{g} Evolution of the central electric flux during the scattering without (green) and with (orange) confinement.
    The experimental data from which these figures were created are given in the Supplementary Information.}
    \label{fig:fermion_scattering}
\end{figure}

\begin{figure*}[t!]
    \centering
    \includegraphics[width=0.95\linewidth]{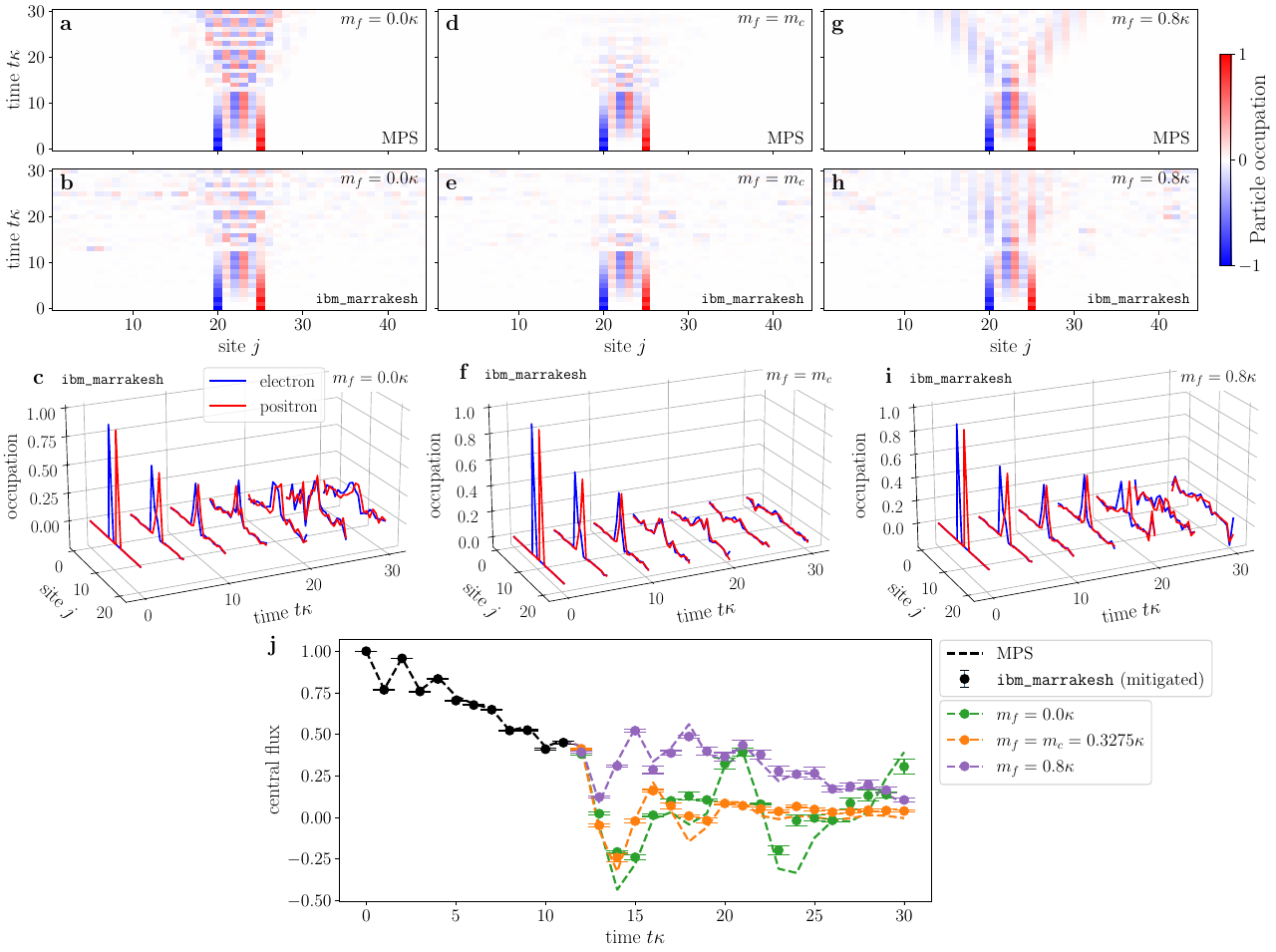}
    \caption{Dynamics of the particle occupation number for electron-positron scatterings with a mass quench at $t=12 \kappa^{-1}$ from the initial mass $m_i = 1.5 \kappa$ to a final mass of $m_f = 0.0 \kappa$ (\textbf{a}-\textbf{c}), $m_f = m_c$ (\textbf{d}-\textbf{f}), and $m_f = 0.8\kappa$ (\textbf{g}-\textbf{i}), respectively. \textbf{a},\textbf{d},\textbf{g} Classical reference calculations performed with MPS. \textbf{b},\textbf{e},\textbf{h} Mitigated results obtained on quantum hardware. \textbf{c},\textbf{f},\textbf{i} Different representation of the same data. Electron occupation number is shown in blue, positron occupation number in red. \textbf{j} Evolution of the central electric flux during the scattering including a mass quench to $m_f = 0.0\kappa$ (green), $m_f = m_c$ (orange), and $m_f = 0.8\kappa$ (purple). The central electric flux before the mass quench is shown in black.
    For all simulations, the chosen value for the confining potential is $\chi = 0.0 \kappa$.
    The experimental data from which these figures were created are given in the Supplementary Information.}
    \label{fig:mass_quenches}
\end{figure*}

After preparing the gauge-invariant initial states, we then develop quantum circuits that realize the gauge-invariant dynamics, which is vital for the study of particle collisions on the quantum computer.
The time evolution is implemented via a first-order Trotter-Suzuki approximation~\cite{hatano2005finding} of the unitary time-evolution operator $\exp\big({-}i t \hat{H}\big)$.
We split the gauge-invariant coupling terms $\hat{P}_{j-1} \hat{\sigma}^x_j \hat{P}_{j+1}$ into two sets of commuting operators, depending on whether $j$ is even or odd. 
The quantum circuit to implement the coupling term $\exp\big({-}i \Delta t \frac{\kappa}{2} \hat{P}_{j-1} \hat{\sigma}^x_j \hat{P}_{j+1} \big)$ is given in Fig.~\ref{fig:quantum_circuit}\textbf{c} and consists of four $R_X$ rotations with angle $\theta = \frac{\kappa}{2} \Delta t / 2$ and four CZ gates.
The time step $\Delta t$ is given by $\Delta t = t / n$ where $t$ is the final time of the simulation and $n$ is the number of Trotter steps.
All commuting even (odd) coupling terms can be applied in parallel, by careful interleaving of the entangling gates.
Implementing the evolution of the coupling terms therefore requires 8 two-qubit gate layers per Trotter step.
The particle mass and the background electric field can be jointly implemented as a single layer of $R_Z$ rotations with a site-dependent angle of $\theta_j = [(-1)^j \chi-2m] \Delta t / 2$.
The full quantum circuit to implement a single first-order Trotter step is displayed in Fig.~\ref{fig:quantum_circuit}\textbf{a}.

For the given initial states, the particles are localized at a single site and each particle will start to delocalize equally in both directions in the subsequent time evolution.
In order to constrain their movement in a desired direction, we remove the coupling terms centered at the neighboring site from the Trotterized quantum circuit. 
This creates a barrier that reflects the component of the wave function moving in its direction and, therefore, effectively creates a wave packet that is moving away from the barrier~\cite{su2024cold}.

We apply a Trotter step of $\Delta t = 1 \kappa^{-1}$ for all executed simulations.
This relatively large step size allows us to explore more of the post-collision dynamics, before reaching the depth limit of the quantum circuits that can reliably be executed on current quantum hardware.
Although such a large Trotter step may lead to deviations from the exact dynamics, we qualitatively still observe all the desired effects in the $\mathrm{U}(1)$ LGT.
The classical reference values we report below were obtained via the simulation of the applied quantum circuits using matrix product states (MPS) \cite{Uli_review,Paeckel_review}, and therefore include the same Trotter errors.

\textbf{Electron-positron scattering.---}We first perform a scattering experiment between an electron and a positron for a system with $45$ sites. We start with a mass of $m=1.5\kappa$, so that vacuum fluctuations are largely suppressed. 
The initial state is an electron-positron pair at a distance of $2$ (staggered) sites.
The walls to constrain the initial propagation of the electron and positron toward each other are included up to Trotter step $10$.
The total time evolution is performed up to $30$ Trotter steps. In the absence of the confining potential ($\chi = 0.0 \kappa$), as shown in Fig.~\ref{fig:fermion_scattering}\textbf{a},\textbf{c},\textbf{e}, the electron-positron pair undergoes an elastic collision, where the electron and positron propagate ballistically away from each other and delocalize after colliding. To better illustrate their trajectory, we run the same quantum circuit with the background vacuum as the initial state, and subtract it from the evolution with particles.
To demonstrate the effect of the confining potential, we then switch to $\chi = 0.15 \kappa$. The confining potential restricts the movement of the electron and positron, where a linear string tension pulls them together after initially propagating away from each other, as shown in Fig.~\ref{fig:fermion_scattering}\textbf{b},\textbf{d},\textbf{e}.
The effect of confinement can also be seen in the central electric flux $\langle \hat{E}_{L/2} \rangle = (-1)^{L/2} \langle \hat{\sigma}^z_{L/2} \rangle$, with $L/2$ being the site at the center of the system; see Fig.~\ref{fig:fermion_scattering}\textbf{g}. Whereas the central flux evolves to a much lower value at the end of the experiment in the case of $\chi=0$, it seems to oscillate around a large steady-state value for $\chi=0.15\kappa$, indicative of an oscillatory bound meson.

All hardware results are reported with central values and statistical error bars as obtained with mDEM. The method itself introduces two hyperparameters: a regularization of the learned noise channel, and a neighborhood size around each local observables (see Supplementary Information). In the present work, we have used hyperparameters of the mDEM that lead to the smallest deviation compared to the reference MPS simulations. 
In future work, we will investigate self-consistent approaches for obtaining suitable hyperparameter values from hardware data only.

In all Figures, the qualitatively different behavior of the post-collision dynamics between $\chi = 0.0 \kappa$ and $\chi = 0.15\kappa$ can be clearly distinguished. Furthermore, the agreement between MPS data and quantum simulations is very good.
As expected, the accuracy of the results decreases with evolution time. 
The decrease is more prominent for the scattering with no confining potential.
This is because confinement limits the dynamics to a few central sites, whereas without confinement the wave packets of the particles not only move away from each other but also start to delocalize.
This requires resolving small values, which is harder for later evolution times since they are affected by more noise.

\begin{figure}[t!]
    \centering
    \includegraphics[width=1.0\linewidth]{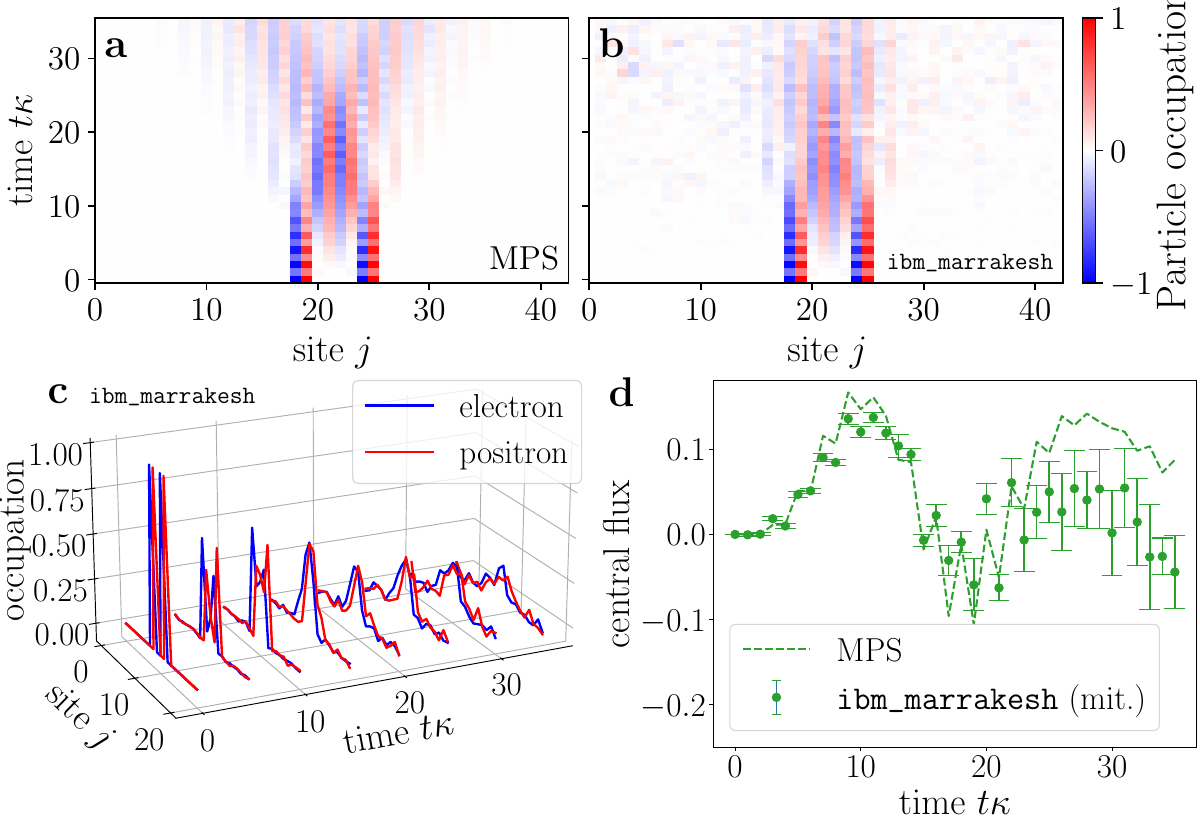}
    \caption{Dynamics of the particle occupation number for a meson-meson collision for $m = 1.5 \kappa$ and $\chi = 0.01 \kappa$. \textbf{a} Classical reference calculation performed with MPS. \textbf{b} Mitigated results obtained on quantum hardware. \textbf{c} Different representation of the data in (b). Electron occupation number is shown in blue, and positron occupation number in red. \textbf{d} Evolution of the central electric flux.
    The experimental data from which these figures were created are given in the Supplementary Information.}
    \label{fig:meson_scattering}
\end{figure}

In addition to tuning the confining potential, interesting physics can also be observed when quenching the mass from an initial value of $m_i = 1.5 \kappa$ to a different final value $m_f$ at the collision time.
In the following, we perform three different quenches of the mass $m_f = 0.0\kappa$, $m_f = m_c = 0.3275\kappa$ and $m_f = 0.8\kappa$, all while $\chi=0$.
We use the same experimental setup as for Fig.~\ref{fig:fermion_scattering}.
The only difference is that for time steps $t \geq 12 \kappa^{-1}$ we use $m_f$ as the value for the mass. The mass quenches lead to prominent meson production in the background, which destabilizes the QED vacuum and results in inelastic collision dynamics.
Figures~\ref{fig:mass_quenches}\textbf{a}-\textbf{i} show the particle occupation numbers, and Fig.~\ref{fig:mass_quenches}\textbf{j} shows the central electric flux, again highlighting impressive agreement between MPS simulations and hardware data. We see three distinct regimes. For $m_f=0$, we enter a regime of inelastic scattering, where we find persistent meson production and annihilation in the post-collision dynamics, which leads to string inversions between the colliding electron and positron \cite{Surace2020}, a phenomenon that is known to occur in the continuum $1+1$D QED model \cite{Hebenstreit2013}. This behavior is closely related to scarring, which is prevalent in this model \cite{Moudgalya_review,Chandran_review,Halimeh2022robust,VanDamme2023anatomy,Hudomal2022,Daniel2023,Desaules2024robust}, and the oscillatory behavior in the flux in Fig.~\ref{fig:mass_quenches}\textbf{j} is a signature of this phenomenon. Another inelastic scattering regime occurs when quenching to the critical point of Coleman's phase transition $m_f=m_c$, where we see a near-complete annihilation of matter at late times. This can also be seen in Fig.~\ref{fig:mass_quenches}\textbf{j} where the flux relaxes near zero, indicating that the system is locally equivalent to a superposition of both vacua. Finally, elastic scattering is recovered when quenching to a final mass above the quantum critical point, $m_f=0.8\kappa$, with dynamics qualitatively similar to that of Fig.~\ref{fig:fermion_scattering}\textbf{a},\textbf{c},\textbf{e}, where the electron and positron collide and then ballistically spread away from each other.

\textbf{Meson-meson scattering.---}We now perform a meson-meson scattering experiment.
To achieve scattering between two mesons, we initialize the system in a state with two mesons placed at a distance.
The initial state is prepared in a system with $43$ sites, and the simulations are executed up to $35$ Trotter steps.
We apply a small confining potential of $\chi = 0.01 \kappa$ to ensure the stability of the mesons.
The rest of the experimental setup is the same as above.
Figure~\ref{fig:meson_scattering}\textbf{a}-\textbf{c} shows the particle occupation number and Fig.~\ref{fig:meson_scattering}\textbf{d} the central electric flux.

The obtained results are in great agreement with MPS for up to $20$ Trotter steps. 
After that point, the noise becomes larger and reduces the accuracy of the obtained dynamics.
This behavior is also reflected in the evolution of the central electric flux.
Compared to the electron-positron scattering results, the collision point in the meson-meson scattering occurs at a later point in time.
Due to this, the post-collision dynamics are subject to more noise in the meson-meson scattering case and, therefore, are captured less accurately.
Nevertheless, these results represent an important first step towards the large-scale simulation of hadron scattering in LGT on a quantum simulator. We are already able to faithfully capture the scattering process up to slightly beyond the collision point.

\textbf{Conclusion and outlook.---}We have performed the first quantum simulation experiment of scattering in a lattice gauge theory, highlighting the potential utility of quantum computers as complementary venues to particle colliders that can provide a first-principles (\textit{ab initio}) probe of the unitary time evolution of scattering processes. Specifically, we have considered a $\mathrm{U}(1)$ lattice gauge theory that serves as a formulation of $1+1$D quantum electrodynamics with discretized gauge fields and then performed digital quantum simulations of electron-positron and meson-meson scattering over a range of tunable parameters on \texttt{IBM}'s \texttt{ibm\_marrakesh} quantum computer. Based on tuning a topological $\Theta$-term, we have demonstrated how a colliding electron and positron spread ballistically away from each other when this term is vanishing, whereas in its presence, they bind back together after an initial post-collision separation. Furthermore, by tuning the mass, we have demonstrated how we can control post-collision particle production, drawing a connection to quantum many-body scarring. Finally, we have also quantum-simulated meson-meson scattering, reaching long enough times on the quantum device to demonstrate the post-collision stability of the mesons in the presence of the topological $\Theta$-term.

All of our experiments have been compared to classical MPS simulations, which show agreement to a high accuracy far beyond the collision point.
This is largely due to the proposed marginal Distribution Error Mitigation, which effectively reduces biases on noisy local observables at the cost of augmented statistical errors.
In this work, we have fixed the number of measurements per experiment at $50'000$, although the statistical errors on the mitigated results can be improved further by increasing this number.
This is especially true for later evolution times, where more shots are required for an accurate characterization of the noise.
With the continued development of tensor network-based circuit compression techniques, such as those involving matrix product operators~\cite{gibbs2024deepcircuitcompressionquantum,le2025riemannianquantumcircuitoptimization,robertson2024tensornetworkenhanceddynamic}, it may become possible to further compress individual Trotter steps in the future. 
This would enable longer evolution times and simulations of larger quantum systems within practical resource limits.
The accuracy of future experiments will be further complemented by the ongoing and future developments of the quantum hardware.

Our experimental results, validated through numerical benchmarks, highlight the validity of our digital quantum simulation scheme as a powerful platform for future explorations of scattering phenomena in more complex gauge theories including in higher spatial dimensions and with non-Abelian gauge groups. Our protocol is scalable and, therefore, stands as a promising candidate for achieving quantum advantage in the simulation of high-energy physics processes.

\medskip

\textit{Note.---}A parallel submission by Z.~Davoudi, C.-C.~Hsieh, and S.~V.~Kadam is to appear in the same \texttt{arXiv} listing as our paper, where they perform a quantum simulation experiment of hadron scattering in a $\mathbb{Z}_2$ LGT on an \texttt{IonQ} quantum computer.

\medskip

\textbf{Acknowledgments.---}We thank Yahui Chai and Karl Jansen for fruitful discussions.
J.S.~and I.T.~were supported by the NCCR MARVEL, a National Centre of Competence in Research, funded by the Swiss National Science Foundation (grant number 205602).
J.J.O.~and J.C.H.~acknowledge funding by the Max Planck Society, the Deutsche Forschungsgemeinschaft (DFG, German Research Foundation) under Germany’s Excellence Strategy – EXC-2111 – 390814868, and the European Research Council (ERC) under the European Union’s Horizon Europe research and innovation program (Grant Agreement No.~101165667)—ERC Starting Grant QuSiGauge.
This work is part of the Quantum Computing for High-Energy Physics (QC4HEP) working group.
J.C.H.~acknowledges the use of IBM Quantum Credits for this work.
IBM, the IBM logo, and ibm.com are trademarks of International Business Machines Corp., registered in many jurisdictions worldwide. Other product and service names might be trademarks of IBM or other companies. The current list of IBM trademarks is available at \href{https://www.ibm.com/legal/copytrade}{this link}.

\section*{Methods}

\textbf{Classical simulations.---}The classical reference values were obtained via MPS~\cite{perez2007matrix,mcculloch2007density} simulations of the applied quantum circuits using the MPS simulator provided by Qiskit~\cite{qiskit2024}.
For the studied system sizes and evolution times, the MPS simulations could be performed within a reasonable amount of time by implementing each gate application up to a fidelity error of $10^{-8}$ and without truncation based on the bond dimension.

\textbf{Quantum simulations.---}We ran all hardware experiments on \texttt{ibm\_marrakesh}, an IBM Quantum device with 156 qubits. 
The active simulation qubits were selected based on the readout and two-qubit gate error rates. 
To evaluate these error rates, we ran benchmarking experiments~\cite{bravyi2021mitigating,magesan2011scalable,magesan2012characterizing} prior to the actual experiments, and chose the line of qubits with the highest value for the product of the corresponding measurement and two-qubit gate fidelities.
Since device properties change over time, we split the workload in batches of 10 experiments (10 physics circuits and 10 noise estimation circuits) and performed the device benchmark for each batch.

During the circuit executions, we applied hardware-level dynamical decoupling to mitigate the effect of cross-talk. Additionally, Pauli twirling~\cite{Bennett1996PurificationNoisyEntanglement} of the entangling CZ gates is implemented with 500 randomizations, and 100 measurements per randomization.
This yields a total number of measurements per estimated expectation value of 50'000.

By default, we set the parameters of the $\mathrm{U}(1)$ LGT to $\kappa = 1.0$, $m = 1.5\kappa$ and $\chi = 0.0 \kappa$.
If different values were used for a simulation, the respective values will be mentioned in the respective places.

The (background-subtracted) particle occupation number used to investigate the electron-positron and meson-meson scattering is defined as
\begin{equation}
   \langle O_i \rangle = \langle O^{\text{wp}}_i \rangle - \langle O^{\text{vac}}_i \rangle \,,
\end{equation}
where
\begin{equation}
   \langle O^{\text{wp}/\text{vac}}_i \rangle = \frac{(-1)^i}{2} \left( \langle Z^{\text{wp}/\text{vac}}_i \rangle + \langle Z^{\text{wp}/\text{vac}}_{i+1} \rangle \right)
\end{equation}
is the occupation for the initial state including the particles and the vacuum initial state, respectively, and $\langle Z_i \rangle$ are the mitigated expectation values of the on-site magnetization.

\textbf{Error mitigation.---}The applied error mitigation strategy is based on the recently proposed Distribution Error Mitigation (DEM)~\cite{gonzales2025quantum}, which is a form of Clifford-based error mitigation~\cite{Czarnik2021ErrorMitigationClifford,Zhang2024CliffordPerturbationApproximationa} but for mitigating sample distributions instead of quantum observables.

The global noise channel of the noisy quantum circuit, the \emph{physics circuit},  after the twirling is approximated by a global Pauli channel at first order in Clifford perturbation theory. The parameters of this global Pauli channel are learned on an additional \textit{noise estimation circuit} with the same gate structure as the physics circuit, but where the single-qubit gates are transformed such that the output of the circuit can be evaluated classically.

Through post-processing of the noise estimation circuit samples as well as the classically calculable expected output, one can mitigate the sample probability distribution of the physics circuit. This error mitigation scheme assumes a first-order Clifford approximation of the physics circuit's full noise channel. However, we find that it yields a remarkable agreement of mitigated local and diagonal observables with their MPS reference values even for large Trotter step sizes and long evolution times. 

The main difference of the applied marginal DEM to the original proposal of DEM is in the post-processing of the measurement output of the noise estimation and physics circuits. This adaption was required to scale the method to the system sizes of the present study. 
Details about both error mitigation strategies can be found in the Supplementary Information.

\clearpage
\onecolumngrid
\begin{center}
    \textbf{\large Supplementary Information for \\``Observation of hadron scattering in a lattice gauge theory on a quantum computer'' }\\[5pt]\end{center}
\setcounter{equation}{0}
\setcounter{figure}{0}
\setcounter{table}{0}
\setcounter{page}{1}
\setcounter{section}{1}
\makeatletter
\renewcommand{\theequation}{S\arabic{equation}}
\renewcommand{\thefigure}{S\arabic{figure}}
\renewcommand{\thesection}{S\Roman{section}}
\renewcommand{\thepage}{\arabic{page}}
\renewcommand{\thetable}{S\arabic{table}}
\vspace{0cm}
\normalsize

\section{Error mitigation}
\label{app:error_mitigation}

In this section, we shortly revisit the Distribution Error Mitigation (DEM) proposed by Gonzales~\cite{gonzales2025quantum-S}. DEM is a Clifford-based technique~\cite{Czarnik2021ErrorMitigationClifford-S,Zhang2024CliffordPerturbationApproximationa-S} that mitigates the effect of a Pauli error channel on the output probability distributions of quantum circuits. While the initial method scales with the support of the noisy quantum state distribution, we present an adaption of the scheme with improved scaling for estimating local diagonal observables.
We additionally show the effectivity of this error mitigation approach on some example experiments from the main text.

Let $\rho$ be a quantum state prepared by a given $n$-qubit quantum circuit, and $\vec{x} \in \mathbb{R}^{2^n}$ be the ideal output probability distribution with $x_i = \Tr(\rho \ketbra{i}{i})$. 
Due to hardware-induced noise, the execution of a quantum circuit effectively results in measurement of noisy samples $\vec{z}$ corresponding to a noisy quantum state $\tilde{\rho}$, with $z_i = \Tr(\tilde{\rho} \ketbra{i}{i})$. 
The ideal and noisy quantum states are generally related by a noise channel $\tilde{\rho} = \varepsilon(\rho)$, which is a completely positive and trace preserving map. 

In DEM, we approximate the global noise channel $\varepsilon$, to lowest order in Clifford perturbation theory, by the noise channel $\varepsilon_{\mathrm{nec}}$ corresponding to the same quantum circuit where all single qubit quantum gates have been replaced by the closest Clifford gate. The resulting quantum circuit, which we will refer to as \emph{noise-estimation circuit}, is Clifford (for our choice of universal gate set). 
As such, the ideal output distribution $\vec{x}_{\text{nec}}$ can be efficiently sampled classically~\cite{Gottesman1998HeisenbergRepresentationQuantum-S}. Execution of the noise-estimation Clifford circuit on hardware results in the noisy sample distribution $\vec{z}_{\text{nec}}$, which, upon twirling of local entangling gates, is related to $\vec{x}_{\text{nec}}$ via a Pauli channel, or equivalently via~\cite{gonzales2025quantum-S}
\begin{equation}
    \vec{z}_\text{nec} = A \vec{x}_\text{nec} \,,
\end{equation}
where $A$ is a $\mathbb{R}^{2^n \times 2^n}$ stochastic matrix (i.e. entries in each row or column sum to 1) with block-circulant structure.
Such a matrix has the useful property of being fully determined by a single column (or row), and that it can be efficiently diagonalized with a Walsh-Hadamard transformation~\cite{rezghi2011diagonalization-S}.
We can fully characterize the block-circulant matrix $A$, or equivalently its first column $\vec{a}$, from the noise characterization circuit only using
\begin{equation}
    \vec{a} = \text{iwfht}\left( \text{fwht}(\vec{z}_\text{nec}) ./ \text{fwht}(\vec{x}_\text{nec}) \right) \,,
\end{equation}
where $./$ is element wise division, $\text{fwht}$ is the Fast-Walsh-Hadamard transform, and $\text{ifwht}$ its inverse.
Once the vector $\vec{a}$ of parameters characterizing the noise channel has been learned on the noise estimation circuit, it can be used to mitigate the noisy distribution $\vec{z}$ of the desired, non-Clifford, quantum state $\tilde \rho$
\begin{equation}\label{eq:dem_inversion}
    \vec{x}_\text{mitig} = \text{iwfht}\left( \text{fwht}(\vec{z}) ./ \text{fwht}(\vec{a}) \right) \,.
\end{equation}
Additionally, we increase the numerical stability of the inversion in Eq.~\eqref{eq:dem_inversion} by introducing a eigenvalue regularization $\epsilon$ that project the learned noise map $A$ to a completely positive map
\begin{equation}\label{eq:truncation}
    \vec{a}_{\text{nec}} = \max(\vec{a}_{\text{nec}}, \epsilon) \,.
\end{equation}

Assuming that the assumptions of DEM are satisfied, diagonal expectation values evaluated from the mitigated distribution $\vec{x}_\text{mitig}$ are estimators of the exact expectation values with reduced bias and increased error bars. The increased statistical error, or equivalently the mitigation overhead, is generally related to the one-norm of the inverse $A^{-1}$ of the noise matrix~\cite{Bravyi2021MitigatingMeasurementErrors-S}. In this work, we instead obtain estimates for the error of the mitigated values of the on-site magnetization $\langle Z_i \rangle$ using bootstrapping.
Given the measured output distributions $\vec{z}$ and $\vec{z}_{\text{nec}}$, we randomly sample the measurement outcomes (with replacement) to obtain the new distributions $\vec{z}^{(k)}$ and $\vec{z}^{(k)}_{\text{nec}}$, each having the same number of measurements as the original ones.
We then use the DEM to evaluate local observables for each mitigated bootstrap sample $\langle Z^{(k)}_i \rangle$. 
We repeat this process 100 times, to get a set of bootstrap samples $\{ \langle Z^{(k)}_i \rangle\}$ from which we then evaluate the average and standard deviation for $\langle Z_i \rangle$.

One main limitation of DEM is that the involved vectors are still of size $2^n$.
To improve the scaling, we propose an adaption of the original approach to marginal distributions, called marginal DEM (mDEM).
Given a set of qubits $C$, we consider the action of extracting marginal distributions restricted to $C$ from all probability distributions on $n$ qubits.
Since this restriction preserves the block-circulant structure of $A$, we can apply the same approach as above to perform the error mitigation on the reduced subspace $C$.
In our case, we perform the experiments on a linear topology, therefore our region of qubits $C$ is a segment of a line.
All properties evaluated in main text are derived from the on-site magnetizations $\langle Z_i \rangle$.
For each site $i$, we define a corresponding set of qubits $C_i = \{ i - n_C, i - n_C + 1, \ldots, i + n_C \}$ of size $2 n_C + 1$.
In this way, the mDEM is fully characterized by two hyperparameters, the regularization parameter $\epsilon$ and the number of considered neighbouring qubits $n_C$ around a single site $i$, which can be tuned in order to obtain the best performance.
For the results presented in the main text, the hyperparameters are selected to minimize the root-mean-square deviation from the MPS simulations
\begin{equation}\label{eq:rmse}
    \text{RMSE}(\{ \langle Z^{\text{Q}}_i \rangle \}, \{ \langle Z^{\text{C}}_i \rangle \}) = \sqrt{\sum_{i = 1}^{n} \left( \langle Z^{\text{Q}}_i \rangle - \langle Z^{\text{C}}_i \rangle \right)^2} \,,
\end{equation}
where $\{ \langle Z^{\text{Q}}_i \rangle \}$ are the mitigated, background-subtracted on-site magnetizations at a given time-step, and $\{ \langle Z^{\text{C}}_i \rangle \}$ the corresponding values obtained with the MPS simulation.
The hyperparameter optimization is performed for each time-step individually.
We empirically found that the best value of the hyperparameters are in the ranges $\epsilon \in [0.01, 0.1]$ and $n_C \in [4, 9]$.
The effect of the hyperparameters $\epsilon$ and $n_C$ on the learned Pauli noise channel will be studied more rigorously in future work.

Figures~\ref{fig:T1_T15_ep}-\ref{fig:T30_T35_mm} illustrates the effect of mDEM on the measured electric flux.
The occupation numbers reported in the main text can be derived from these data as described in the method section. 

\section{Scattering experiments}

\begin{figure*}
    \centering
    \includegraphics[width=1.0\linewidth]{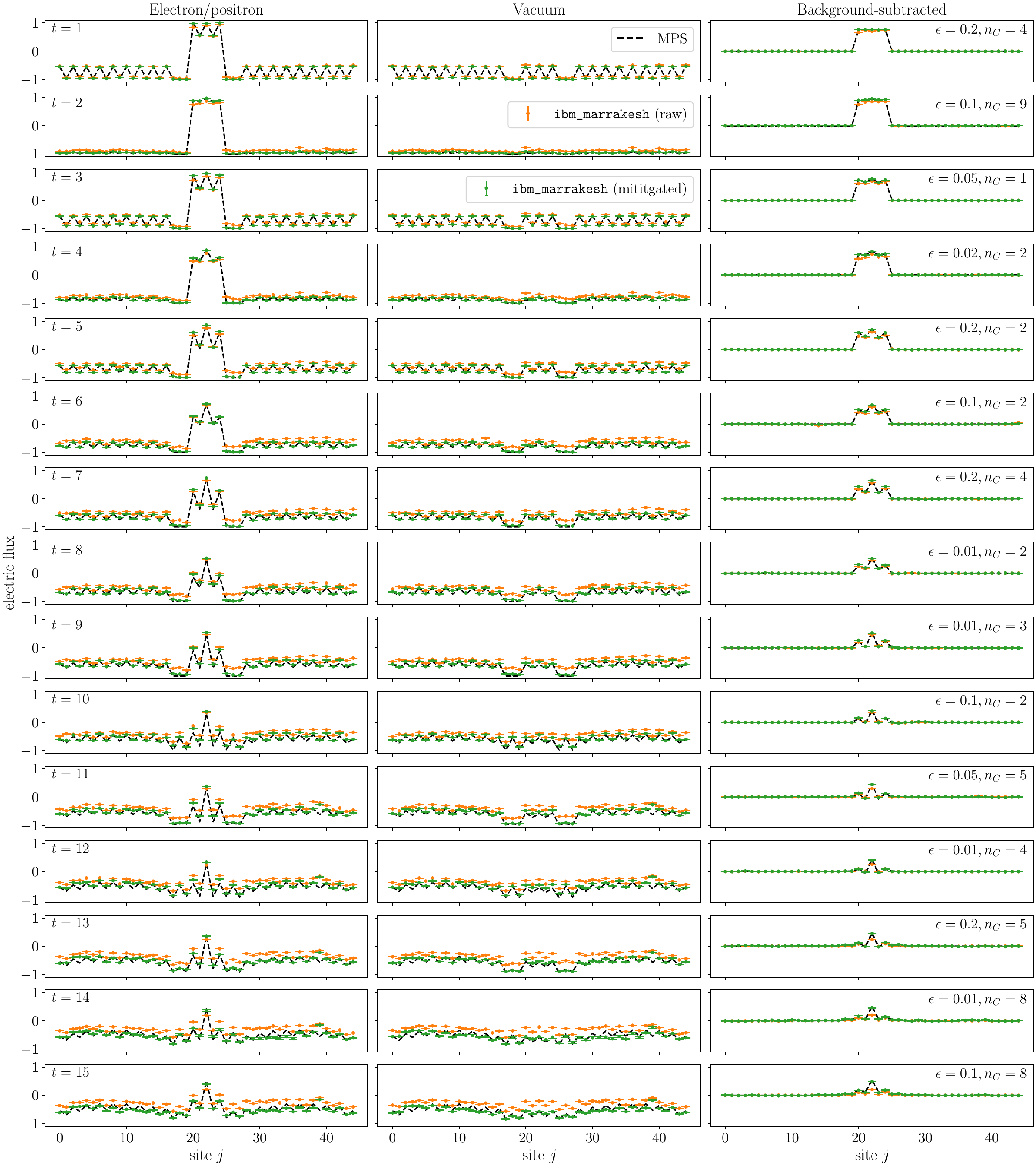}
    \caption{Measured (orange) and mitigated (green) electric flux for the electron-positron scattering without confining potential. Time steps 1 to 15. Classical reference values (black) obtained with MPS simulations.}
    \label{fig:T1_T15_ep}
\end{figure*}

\begin{figure*}
    \centering
    \includegraphics[width=1.0\linewidth]{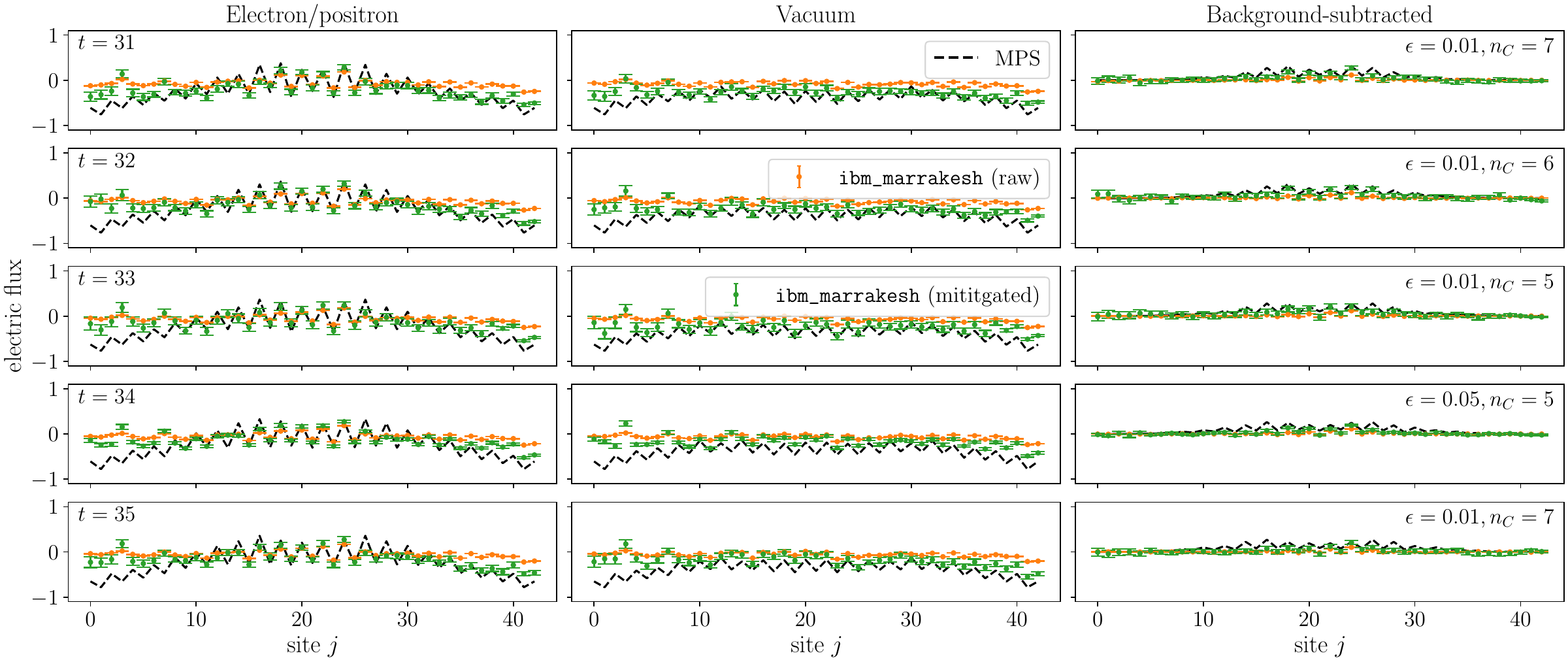}
    \caption{Measured (orange) and mitigated (green) electric flux for the meson-meson scattering. Time steps 30 to 35. Classical reference values (black) obtained with MPS simulations.}
    \label{fig:T30_T35_mm}
\end{figure*}

In the following, we display all experimental data obtained with \texttt{ibm\_marrakesh}.
Figures~\ref{fig:T1_T15_ep}-\ref{fig:T30_T35_mm} show the measured (orange) and mitigated (green) evolution of the electric flux for the wave-packet (left column), the vacuum (middle column), and the background-subtracted electric flux (right column).
The last column also reports the values of the hyperparameters used for the mitigation at the given Trotter step.
The error bars of the raw hardware data correspond to the standard deviation in the underlying measurement data.
The error bars of the mitigated results are obtained with the bootstrapping procedure described above.
Additionally, in table~\ref{tab:quantum_circuits} we report the size of the quantum circuits used in the experiments, for a representative number of Trotter steps.

\begin{table}[ht!]
    \centering
    \begin{tabular}{l | c | c | c | c}
        Experiment & Num. qubits & Num. Trotter steps & Num. 2q gates & 2q gate depth \\
        \hline
        electron/positron & 45 & 10 & 1688 & 80 \\
                          &    & 20 & 3448 & 160 \\
                          &    & 30 & 5208 & 240 \\
        \hline
        meson/meson & 43 & 10 & 1608 & 80 \\
                    &    & 20 & 3288 & 160 \\
                    &    & 30 & 4968 & 240 \\
                    &    & 35 & 5808 & 280 \\
    \end{tabular}
    \caption{Size of the applied quantum circuits for certain number of Trotter steps.}
    \label{tab:quantum_circuits}
\end{table}

\begin{figure*}
    \centering
    \includegraphics[width=1.0\linewidth]{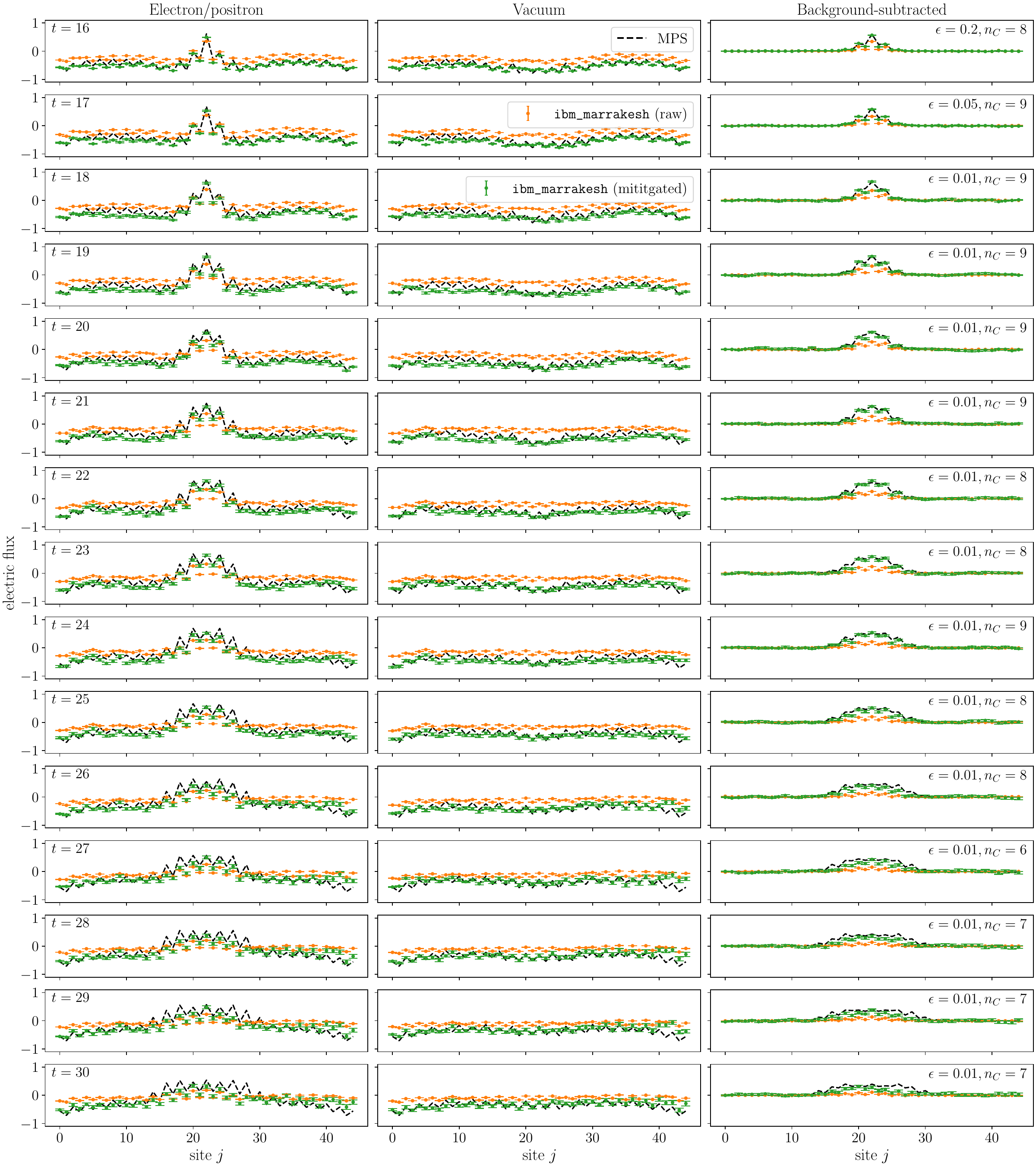}
    \caption{Measured (orange) and mitigated (green) electric flux for the electron-positron scattering without confining potential. Time steps 16 to 30. Classical reference values (black) obtained with MPS simulations.}
\end{figure*}

\begin{figure*}
    \centering
    \includegraphics[width=1.0\linewidth]{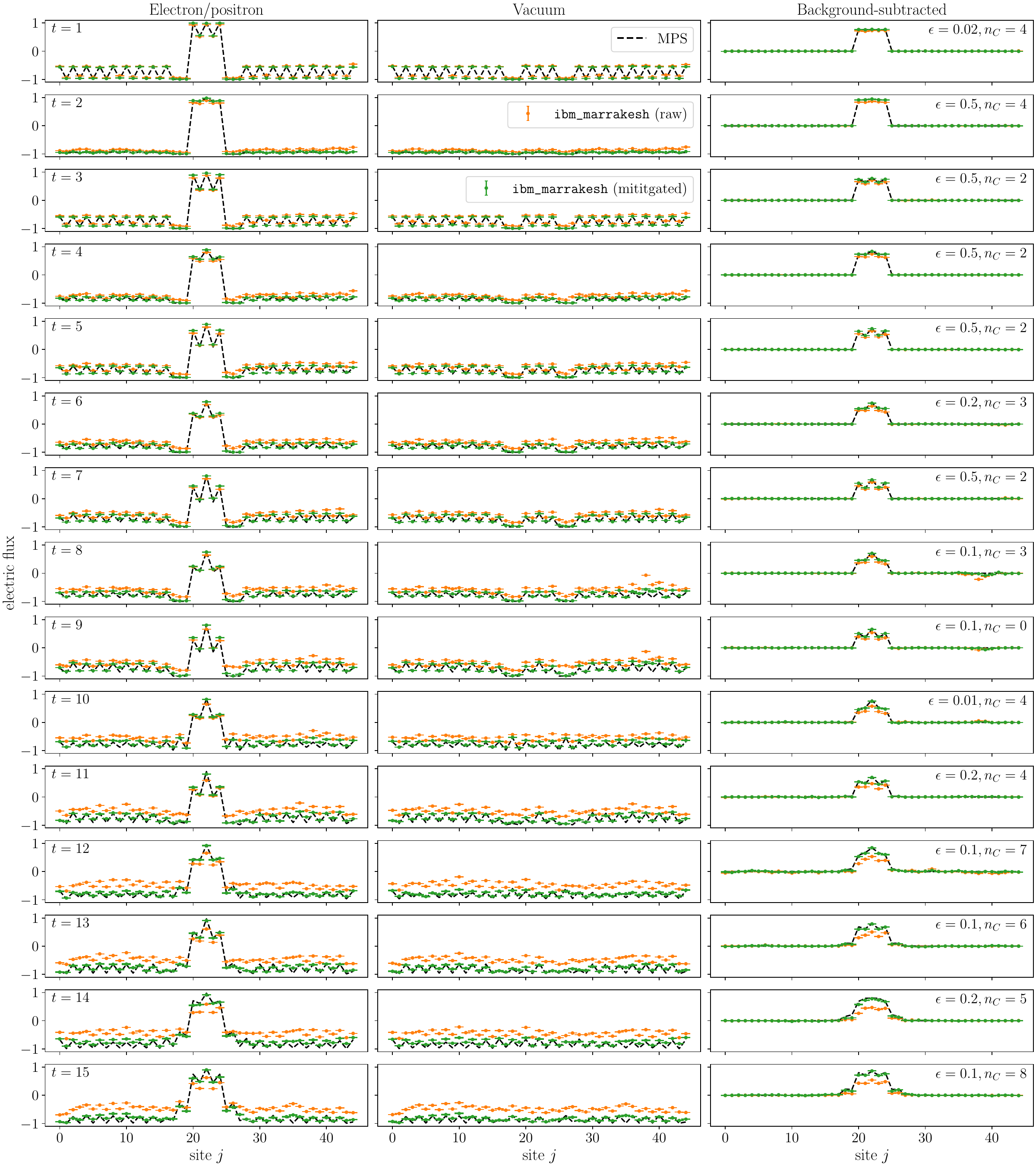}
    \caption{Measured (orange) and mitigated (green) electric flux for the electron-positron scattering with $\chi = 0.15 \kappa$. Time steps 1 to 15. Classical reference values (black) obtained with MPS simulations.}
\end{figure*}

\begin{figure*}
    \centering
    \includegraphics[width=1.0\linewidth]{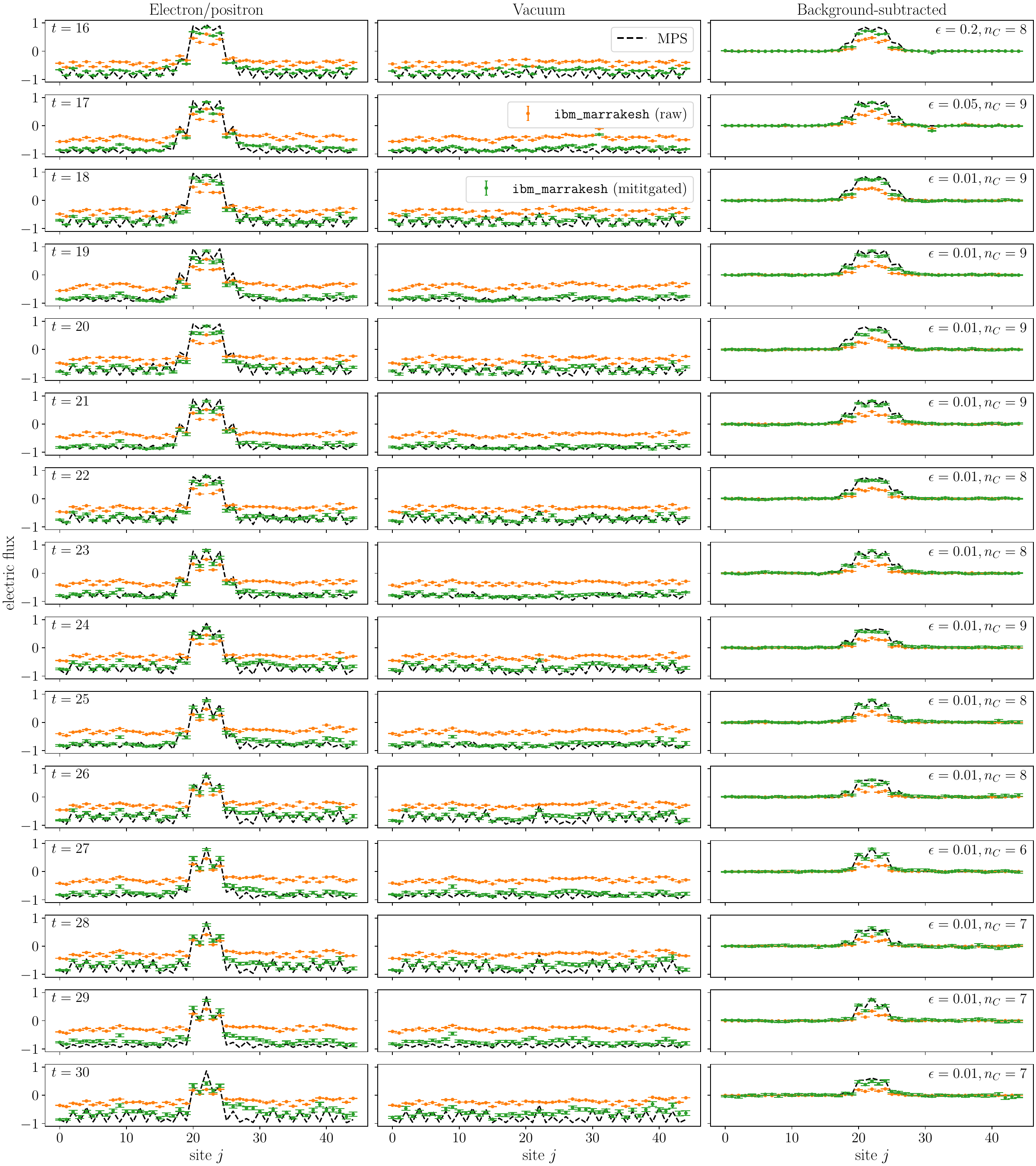}
    \caption{Measured (orange) and mitigated (green) electric flux for the electron-positron scattering with $\chi = 0.15 \kappa$. Time steps 16 to 30. Classical reference values (black) obtained with MPS simulations.}
\end{figure*}

\begin{figure*}
    \centering
    \includegraphics[width=1.0\linewidth]{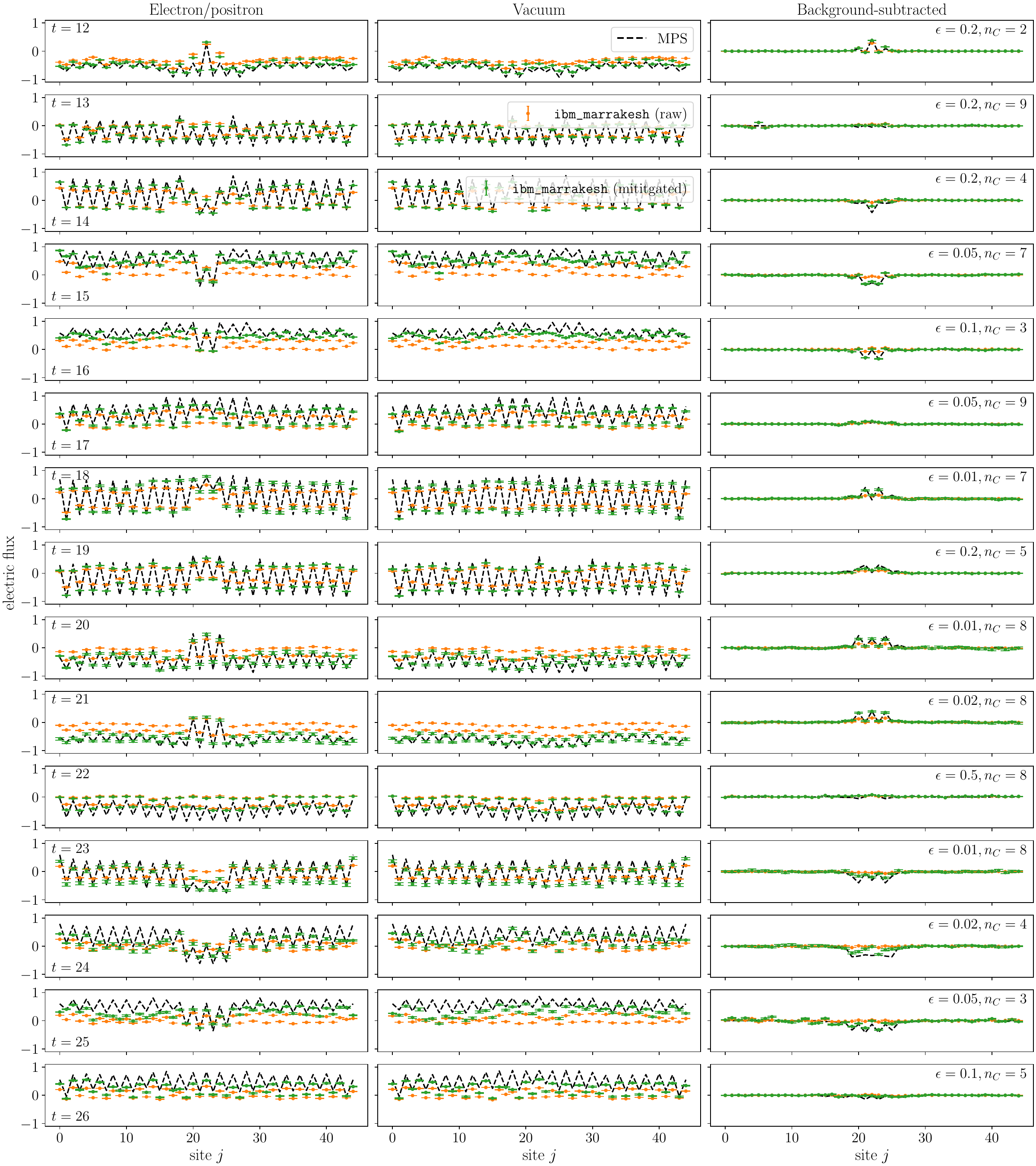}
    \caption{Measured (orange) and mitigated (green) electric flux for the electron-positron scattering after the mass quench to $m_f = 0.0\kappa$. Time steps 12 to 26. Classical reference values (black) obtained with MPS simulations.}
\end{figure*}

\begin{figure*}
    \centering
    \includegraphics[width=0.97\linewidth]{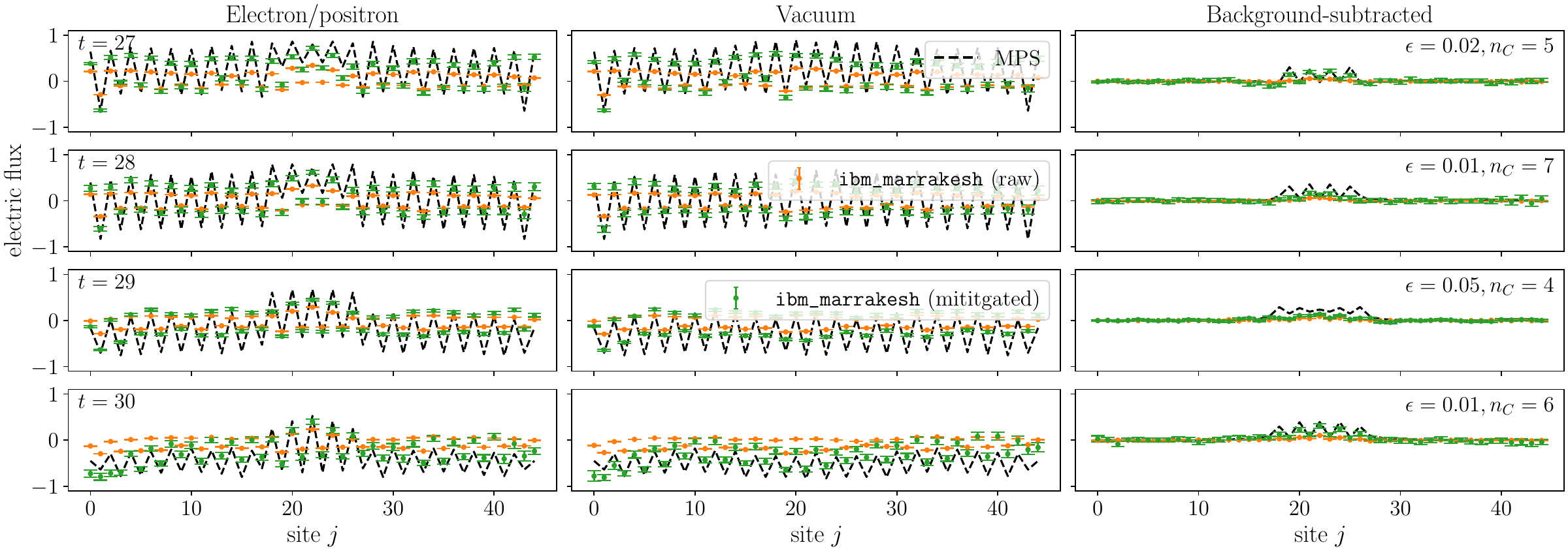}
    \caption{Measured (orange) and mitigated (green) electric flux for the electron-positron scattering after the mass quench to $m_f = 0.0\kappa$. Time steps 27 to 30. Classical reference values (black) obtained with MPS simulations.}
\end{figure*}

\begin{figure*}
    \centering
    \includegraphics[width=0.97\linewidth]{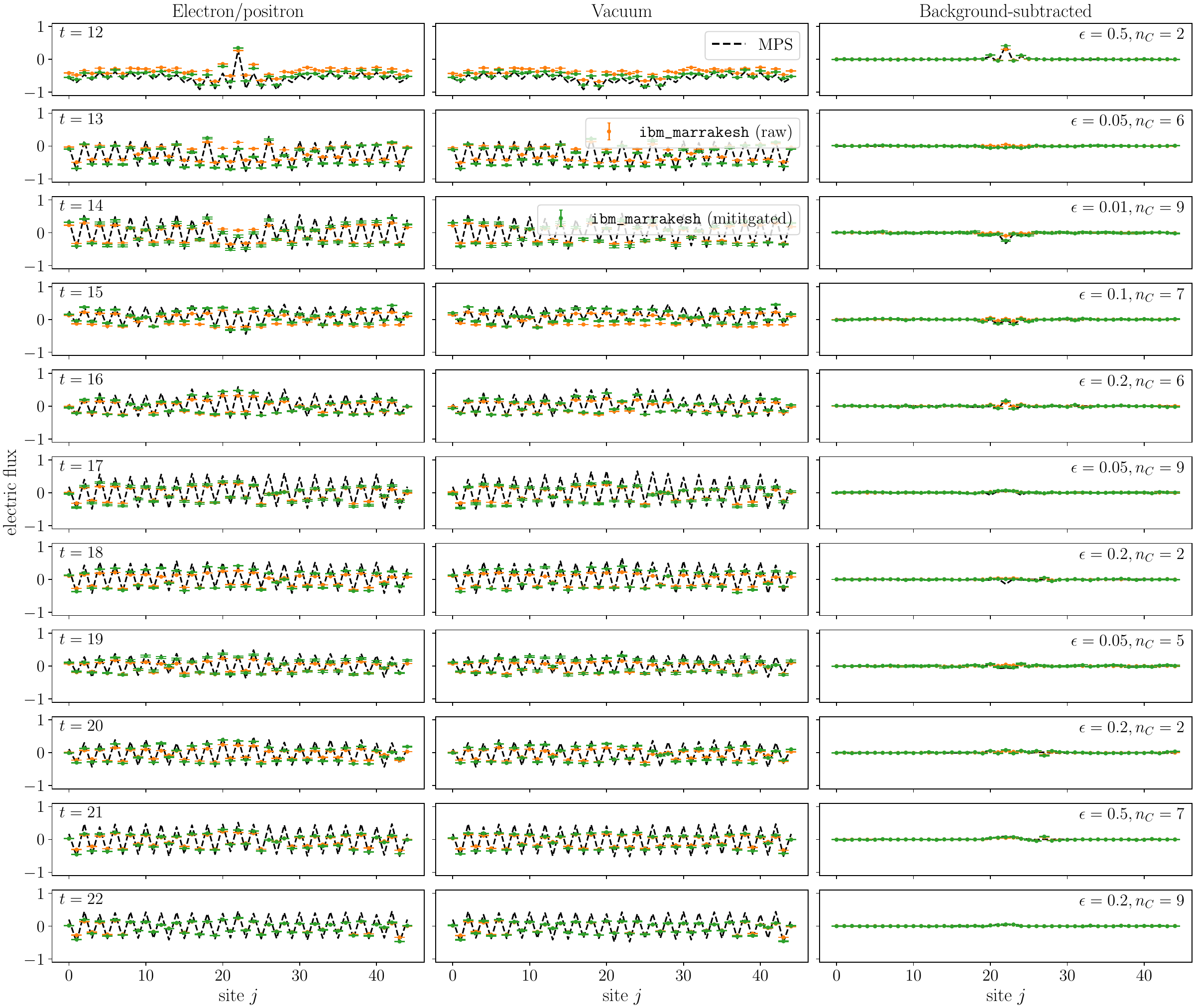}
    \caption{Measured (orange) and mitigated (green) electric flux for the electron-positron scattering after the mass quench to $m_f = m_c = 0.3275\kappa$. Time steps 12 to 22. Classical reference values (black) obtained with MPS simulations.}
\end{figure*}

\begin{figure*}
    \centering
    \includegraphics[width=0.97\linewidth]{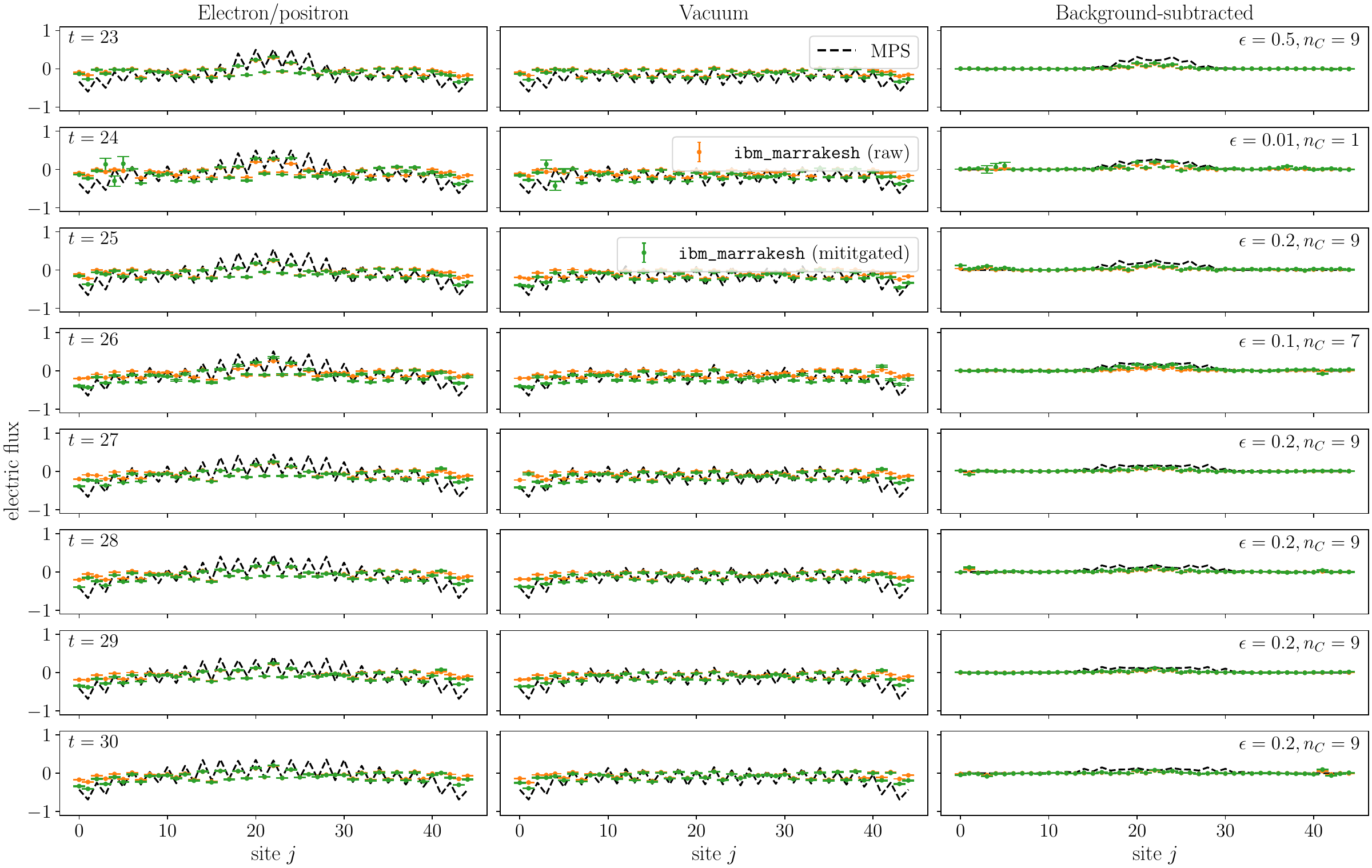}
    \caption{Measured (orange) and mitigated (green) electric flux for the electron-positron scattering after the mass quench to $m_f = m_c = 0.3275\kappa$. Time steps 23 to 30. Classical reference values (black) obtained with MPS simulations.}
\end{figure*}
\begin{figure*}
    \centering
    \includegraphics[width=0.97\linewidth]{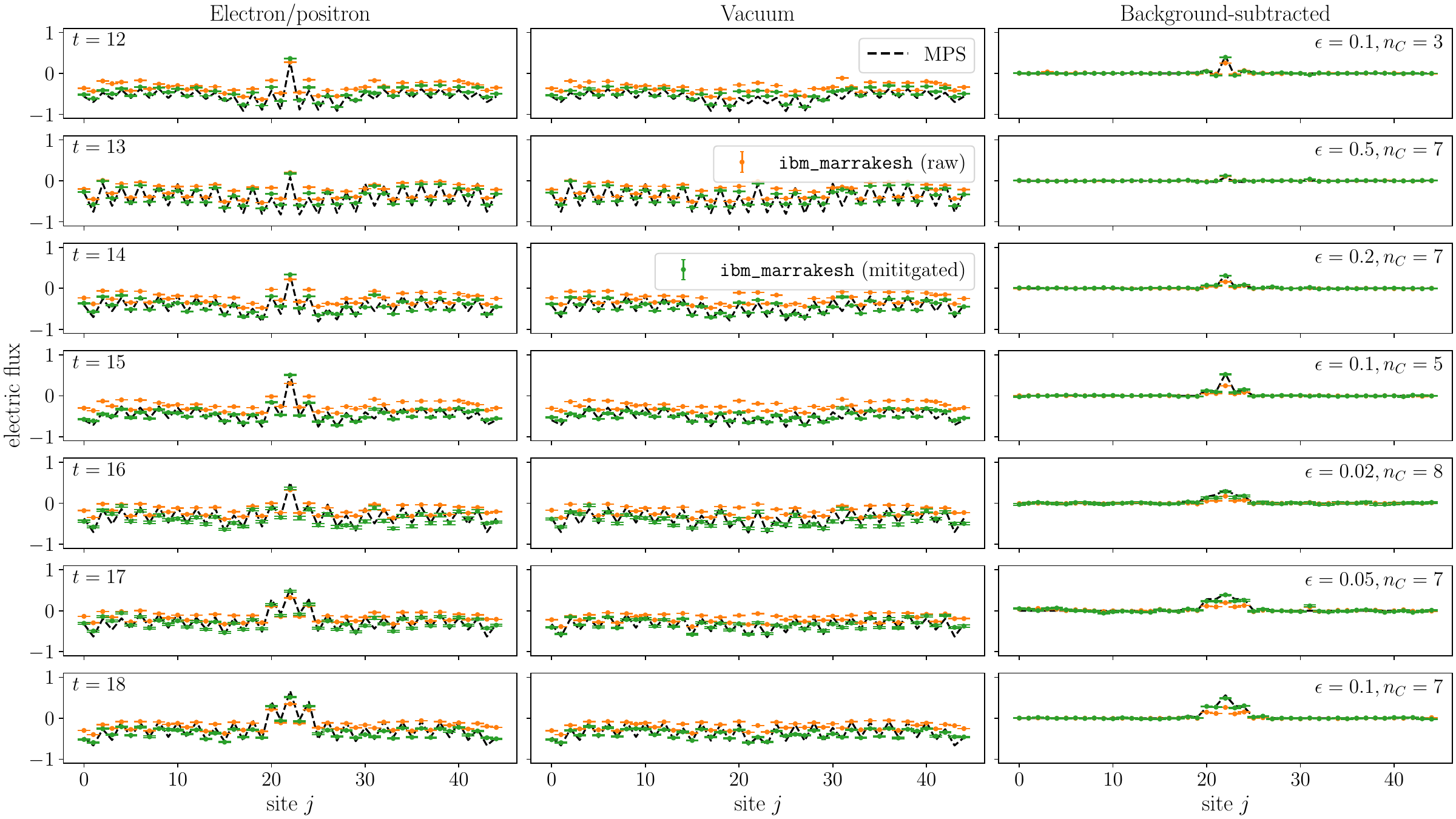}
    \caption{Measured (orange) and mitigated (green) electric flux for the electron-positron scattering after the mass quench to $m_f = 0.8\kappa$. Time steps 12 to 18. Classical reference values (black) obtained with MPS simulations.}
\end{figure*}

\begin{figure*}
    \centering
    \includegraphics[width=1.0\linewidth]{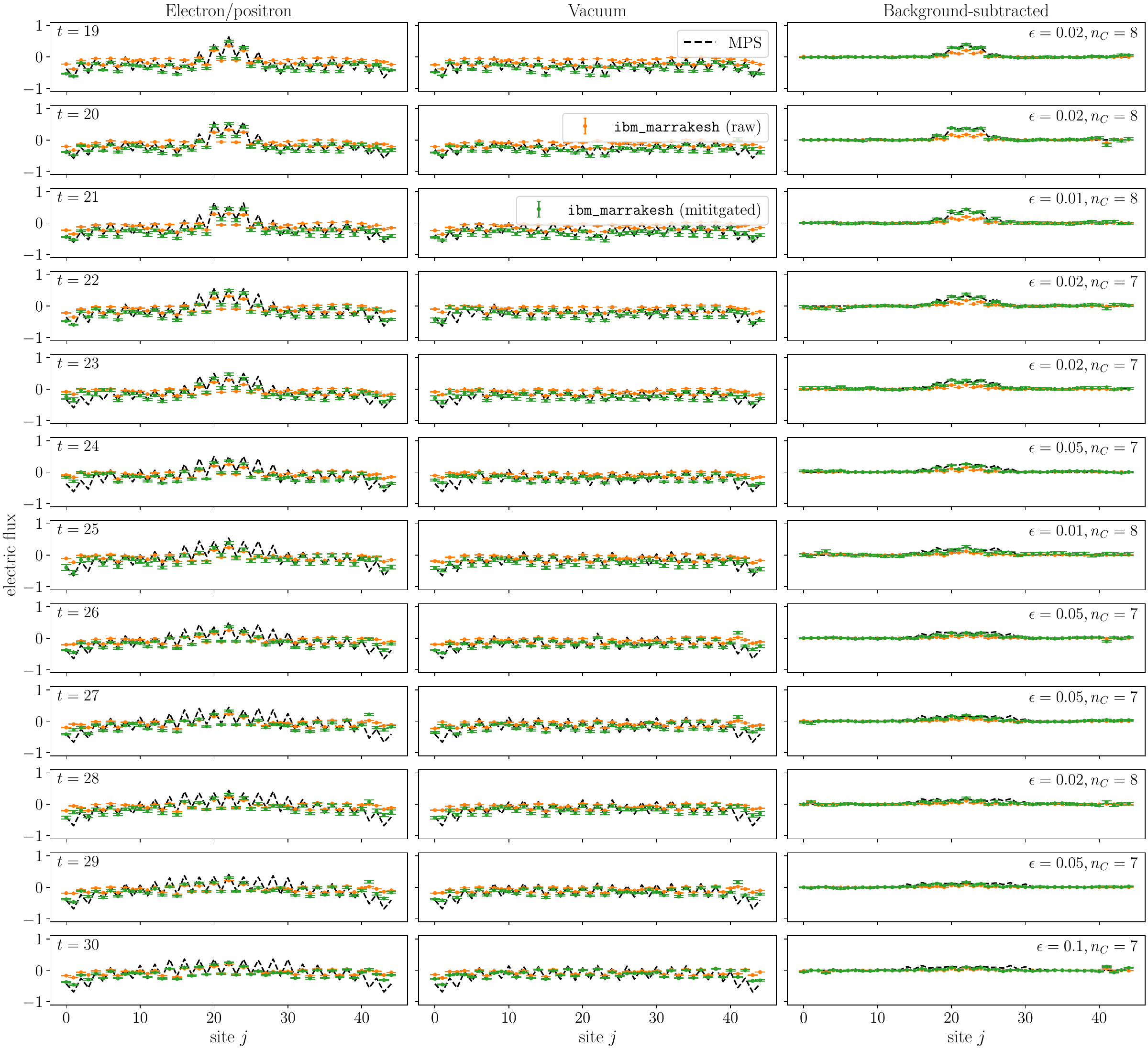}
    \caption{Measured (orange) and mitigated (green) electric flux for the electron-positron scattering after the mass quench to $m_f = 0.8\kappa$. Time steps 19 to 30. Classical reference values (black) obtained with MPS simulations.}
\end{figure*}

\begin{figure*}
    \centering
    \includegraphics[width=1.0\linewidth]{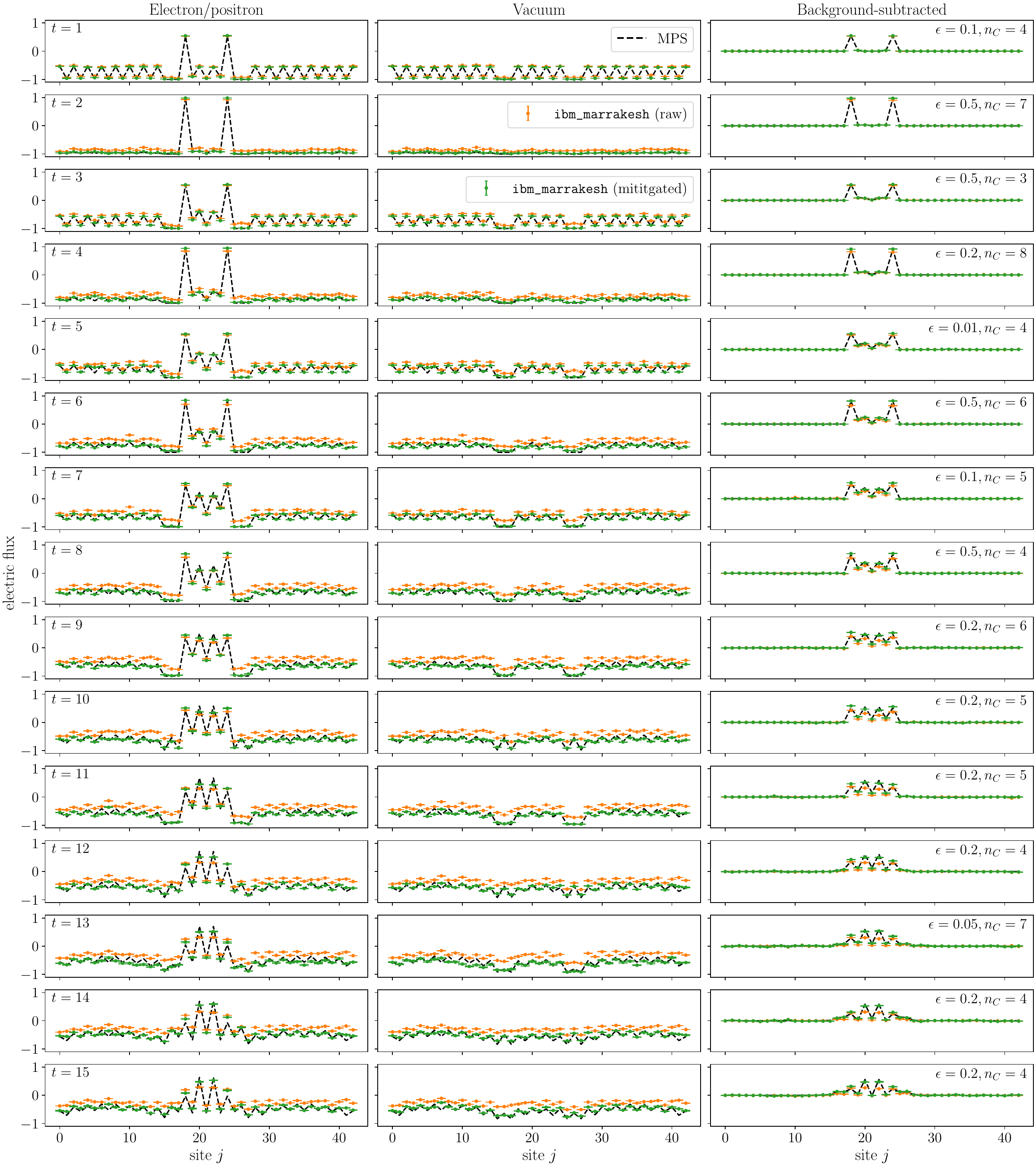}
    \caption{Measured (orange) and mitigated (green) electric flux for the meson-meson scattering. Time steps 1 to 15. Classical reference values (black) obtained with MPS simulations.}
\end{figure*}

\begin{figure*}
    \centering
    \includegraphics[width=1.0\linewidth]{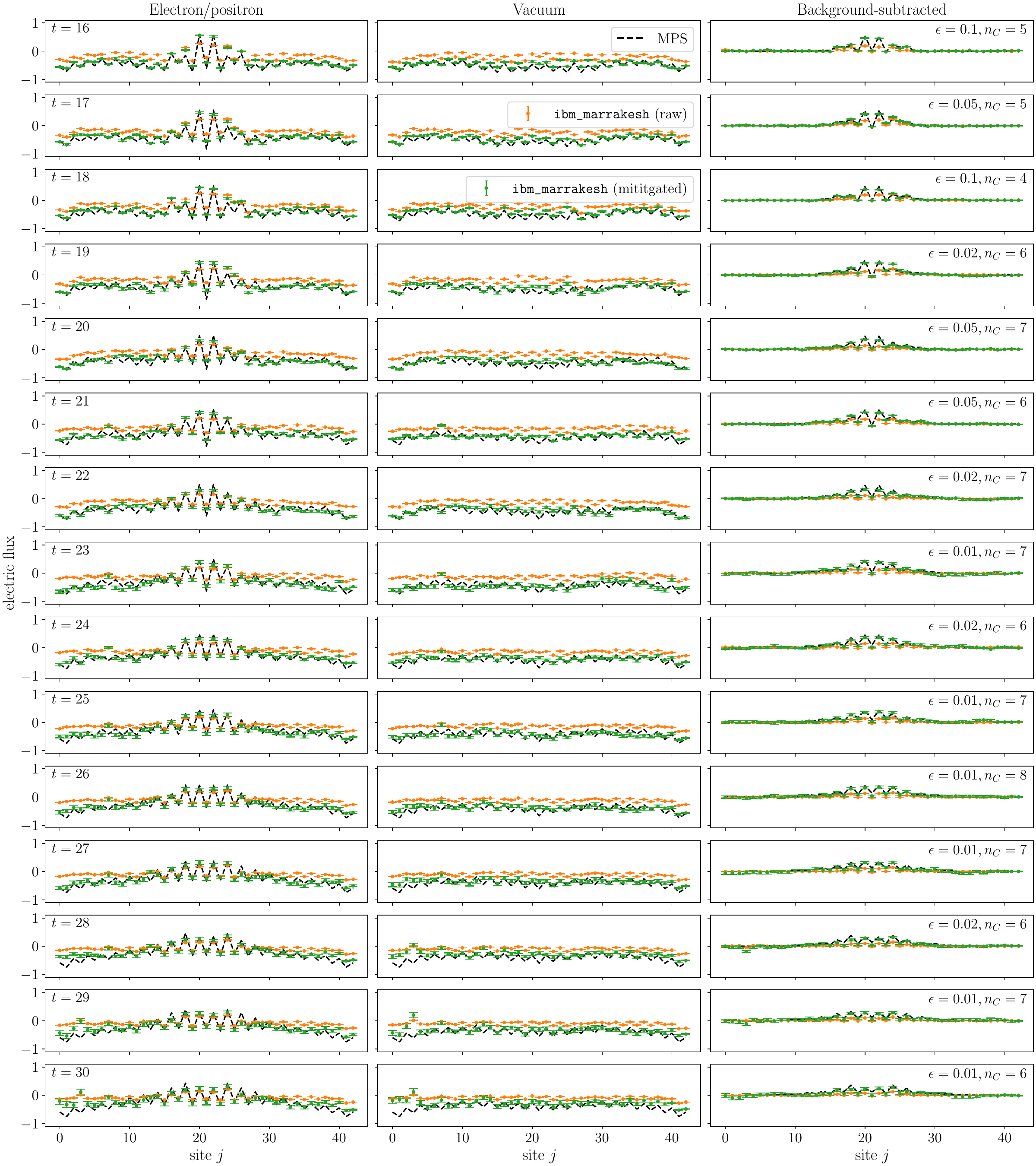}
    \caption{Measured (orange) and mitigated (green) electric flux for the meson-meson scattering. Time steps 16 to 30. Classical reference values (black) obtained with MPS simulations.}
\end{figure*}

\end{document}